\documentclass[twocolumn]{aa}  

\usepackage{graphicx}
\usepackage{txfonts}
\usepackage[sort&compress]{natbib}
\bibpunct{(}{)}{;}{a}{}{,}
\newcommand{\um}{$\mu$m}

\def\degr{\hbox{$^\circ$}}

\def\arcsec{\hbox{$^{\prime\prime}$}}
\def\utw{\smash{\rlap{\lower5pt\hbox{$\sim$}}}}
\def\udtw{\smash{\rlap{\lower6pt\hbox{$\approx$}}}}

\def\Mdot{\hbox{$\dot {M}$}}

\def\Lsun{\hbox{\it L$_\odot$}}

\def\Msun{\hbox{\it M$_\odot$}}

\newcommand{\Ks}{{\it K$_{\rm s}$}}

\newcommand{\Aks}{{\it A$_{\rm K_{\rm s}}$}}
\newcommand{\Ak}{{\it A$_{\rm K}$}}

\def\BCKs{\hbox{\it BC$_{\rm K_S}$}}

\def\BCA{\hbox{\it BC$_{\rm A}$}}
\def\simgr{\mathrel{\hbox{\rlap{\hbox{\lower4pt\hbox{$\sim$}}}\hbox{$>$}}}}

\begin{document}
   \title{Near- and Mid-Infrared colors of  evolved stars in the Galactic plane. 
The $Q1$ and $Q2$ parameters.}

   \subtitle{}

   \author{M. Messineo
          \inst{1}
          \and
          K.~M. Menten
	  \inst{1}
	  and
	  E. Churchwell 
	  \inst{3}
	  and
	  H. Habing
	  \inst{2}
         }

   \institute{Max-Planck-Institut f\"ur Radioastronomie,
Auf dem H\"ugel 69, D-53121 Bonn, Germany\\
              \email{messineo@mpifr-bonn.mpg.de}
         \and
Leiden Observatory, PO Box 9513, 2300 RA Leiden, The Netherlands \\	      
         \and
Department of Astronomy, University of Wisconsin-Madison, 475 N. Charter street, Madison, WI 53706, USA	 
             }


   \date{Received September 15, 1996; accepted March 16, 1997}

 
  \abstract
   {Mass-loss from evolved stars chemically enriches the interstellar medium (ISM). Stellar winds
from massive stars and their explosions as supernovae  shape the ISM 
and  trigger star formation. Studying evolved stars is  fundamental for understanding
 galaxy formation and evolution, at any redshift. 
}
   {We aim to establish a photometric identification and classification scheme for Galactic mass-losing 
evolved stars (e.g., WR, RSG, and AGB stars) with the goal of identifying new ones, and
subsequently to use the sample as tracers  of Galactic structure.
}
   {We  searched for counterparts of known Galactic  WR, LBV, RSG, and O-rich AGB stars   
   in the 2MASS, GLIMPSE, and MSX catalogs, and we  analyzed their properties with near- 
   and mid-infrared color-color diagrams.
}
   {We used the $Q1$  parameter,  which is a measure of  the deviation from the
interstellar reddening vector in the $J-H$ versus $H-$\Ks\ diagram, and  we defined 
a new parameter, $Q2$,  that is a measure of the deviation from the interstellar reddening vector in the
$J-$\Ks\ versus \Ks$-[8.0]$ diagram. The latter plane enables 
to distinguish between interstellar and circumstellar reddening, and to identify  stars with 
circumstellar envelopes. 
WR stars  and late-type mass-losing stars (AGBs and RSGs) are distributed in two different regions
of the $Q1$ versus \Ks$-[8.0]$ diagram. A sequence of increasing $[3.6]-[4.5]$ and 
$[3.6]-[8.0]$ colors with  increasing  pulsation amplitudes (SRs, Miras, and OH/IR stars) is found. 
Spectra of Miras and OH/IR stars have stronger water absorption at 3.0 \um\ than SR stars or most of the RSGs.
Masing Miras stars   have water, but stronger SiO ($\sim 4$ um) and CO$_2$ absorption ($\sim 4.25$ \um), as
suggested by  their $[3.6]-[4.5]$ colors, bluer than those of non masing stars.
A fraction of RSGs (22\%) have the bluest $[3.6]-[4.5]$ colors, but small $Q2$ values. 
We propose a new set of photometric criteria to distinguish among  IR bright  Galactic stars. 
}
   {The GLIMPSE catalog is a powerful tool for photometric classification of Galactic
 mass-losing evolved stars. Our new criteria will yield many new RSGs and WRs. 
}


   \maketitle
%

\section{Introduction}

Evolved stars are intrinsically bright and,    when their luminosities can be established,
they  may serve as distance indicators. Thus, they are ideal tracers of Galactic structure. 
Moreover, since they  lose mass at high rates,  they chemically enrich the Galactic
interstellar medium (ISM). 

Intermediate mass stars  (from $\sim 2.2$ to $\sim 8$ \Msun) lose  mass at high rates when
they go trough the Asymptotic Giant Branch phase (AGB stars), and,  thereafter, the planetary
nebula (PN) phase, before cooling as  white dwarfs.  Evolved massive stars ($> 8$ \Msun) go
through  several short-lived evolutionary phases (blue supergiants, BSGs, luminous blue
variables, LBVs,  Wolf-Rayet stars, WRs, yellow supergiants, YSGs, and red supergiants, RSGs).
Eventually, they explode as  supernovae leaving a  neutron star or a black-hole.  Late-type
stars (red giants, AGBs, and RSGs) are  the most prolific dust producers in  the Universe
\citep{gehrz89}.   The archetype VY CMa has a mass-loss rate of 10$^{-4}$ M$_\odot$ $y^{-1}$
\citep[][]{harwit01}. While RSGs may have higher mass-loss, since $\dot{M} \propto L$, 
luminous red giants and AGB  are more numerous and, account for more than 70\% of the dust  in
the interstellar medium. $\dot{M}$ and $L$ are the stellar mass loss rate and luminosity,
respectively  \citep[e.g.][]{gehrz89}. In contrast, the majority of hot massive stars are not
associated  with dusty envelopes;  WRs  produce only 5\% of  the dust  \citep{gehrz89}. While
massive stars are typically found in or near HII regions and well trace galactic spiral arms
\citep{georgelin76}.  AGBs, which  are older and dynamically relaxed, may serve  as direct
tracers of the Galactic gravitational potential  \citep{habing06}.  Kinematics of AGBs  have
been successfully used to map mass components  devoid of gas  \citep[e.g,  galactic
bars][]{habing06, messineo02}. 

A census and classification of Galactic evolved stars is of primary  importance to understand
galaxy formation and evolution.  AGBs and RSGs are still poorly modeled, and  their relative
contribution to the integrated light of a galaxy is quite uncertain. Major difficulties arise
from our poor understanding of the complex process of mass-loss. Moreover, observational
identification of evolved stars is  a difficult  task.   Optical studies cannot penetrate the
inner Galactic regions due to interstellar extinction;  stellar classification in the
near-infrared is more difficult, because intrinsic near-infrared colors of evolved stars span
a small range  \citep{koornneef83}. Spectroscopic observations may solve the ambiguity, but
they requires more telescope time than photometry, and are feasible only for a limited number
of  objects. Photometric variability, mass-loss, interstellar extinction, and poor  knowledge
on stellar distances hamper the detection and classification  even of these brightest stars. 

Several near-infrared and mid-infrared surveys of the whole Galactic plane are  now (or will
soon become) available, e.g. the Two Micron All Sky survey (2MASS)  \citep{cutri03}, the Deep
Near-Infrared Survey (DENIS)  \citep{denis,denis2},  the UKIRT Infrared Deep Sky Survey
(UKIDSS) \citep{lucas08}, the VISTA survey \citep{vista10},  the Midcourse Space Experiment
(MSX) \citep{price01}, the ISO infrared survey of the  Galactic Plane (ISOGAL)
\citep{schuller03}, and the Galactic Legacy Infrared Mid--Plane  Survey Extraordinaire
(GLIMPSE) \citep{benjamin03,churchwell09}.  The combined use of near- and  mid-infrared
measurements permits detection of evolved mass-losing stars, as demonstrated with ISOGAL and
MSX data \citep[e.g.][]{alard01, messineo05}.  However, the classification of evolved stars
using GLIMPSE data is still largely unexplored.  Similar studies have been carried out  on the
Magellanic Clouds   \citep[e.g.][]{yang11,buchanan06,buchanan09,bonanos10}. However, these
studies  suffer  from small number  statistics, and  are limited to  an environment with low
metallicity.

In the Milky Way, we know about 500 RSGs,  350 WR stars \citep{vanderhucht01}, and a dozen
confirmed LBVs \citep{clark05}. Several thousands of AGBs have been  detected via their maser
emission, or photometric pulsation properties   \citep[e.g.][ and references
therein]{habing06,alard01,glass01,deguchi04,messineo02,sevenster02}.  These numbers are too
small for quantitatively constraining theories of Galactic formation  and evolution
\citep[e.g.][]{vauterin98,habing06}, and for significantly sampling short-lived  evolutionary
phases and their contribution to Galactic chemical enrichment.  Galactic models predict  about
9 million Miras, $\sim 5000$ M supergiants,  and $\sim 3000$ WR stars \citep[e.g.][]{gehrz89}.

Stellar classification  based on infrared two-color diagrams  was successfully established in
the late 1980's, based on data from the Infrared Astronomical Satellite Point Source 
catalogue (the IRAS PSC) \citep[e.g.][]{vanderveen88}. The 60\um/25\um\ versus 25\um/12\um\
color-color diagram distinguishes  between a carbon and an oxygen-rich chemistry-, and  shows
a  sequence of O-rich circumstellar shells with  increasing mass-loss rates. Successful
selection of mass-losing late-type stars has  been  made with   data from the MSX survey
\citep[e.g.][]{messineo04,messineo05,sevenster02,sjouwerman09}.  A dust sequence  of O-rich
envelopes is seen also with MSX colors \citep{sjouwerman09, sevenster99}. MSX mapped the whole
Galactic Plane within 5\degr\ of latitude (plus the IRAS gaps)   in six bands  (the $B1$-band
is centered at 4.3 \um, the $B2$ band at 4.35 \um,   the $A$-band at 8.28 \um, the $C$-band at
12.13 \um, the $D$-band at 14.65 \um,  and the $E$-band at 21.34 \um). The $A$-band has  a
sensitivity of 0.1 Jy, and  a spatial resolution of 18.3\arcsec\  \citep{price01}.  We
indicate with [A],[C],[D], and [E], the magnitudes in the 4 MSX bands at 8.28 \um, 12.13\um,
14.65\um,  and 21.34\um, respectively.

The IRAC camera on board the Spitzer Space Telescope  offers a new view of the sky  in four
filters  \citep[bands are centered at 3.6, 4.5, 5.8, and 8.0 \um~][]{fazio04},   with a
spatial resolution 6 times better than MSX, and a better sensitivity.  We indicate with
$[3.6]$, $[4.5]$, $[5.8]$, and $[8.0]$, the magnitudes in the four GLIMPSE filters.
\citet{buchanan09} and \citet{bonanos10} developed a set of color criteria  for classifying 
luminous 8 \um\ sources in the  Large Magellanic Cloud (LMC). HII regions can  be isolated  by
their much redder \Ks$-A$ colors, while   measurements in the four IRAC bands  distinguish
between RSGs,    O-rich  and C-rich AGBs.  \citet{yang11}  revised the pulsation properties of
RSGs in the LMC using   a comprehensive list of 191 RSGs  \citep{buchanan09}.   Only 24\% of
RSGs  show regular pulsation \citep{yang11}. The IR selected sub-sample of  LMC RSGs   appears
to have redder \Ks$-[8.0]$ colors (likely due to higher mass-loss rates) than the  average
sample.  RSGs are found to have bluer $[3.6]-[4.5]$ colors than other luminous red stars. 

So far, little work has been done on Galactic evolved stars based on  GLIMPSE data. WRs have
an infrared excess,  due to  free-free emission from dense stellar winds  \citep{cohen75}.
\citet{hadfield07}  established a successful selection of WRs  based on a combination of 
2MASS and GLIMPSE colors, which has been recently revised by \citet{mauerhan11}. WRs can   be
identified by   combining three color-color diagrams,  the $J-H$ versus $J-$\Ks\ diagram, the
$[3.6]-[8.0]$  versus $[3.6]-[4.5]$ diagram, and the \Ks$-[8.0]$ versus $J-$\Ks\ diagram
(Figs.\   \ref{jhk}, \ref{lucy1} and \ref{lucy2}). Known WRs have  $[3.6]-[8.0] > 0.5$ mag and
$[3.6]-[4.5] >0.1$ mag,  and in the \Ks$-[8.0]$ versus $J-$\Ks\ diagrams are located on a
different sequence than normal OB stars; their distribution  with redder colors does not
follow the reddening vector.    Hadfield et al. reported a detection rate of 65\% for emission
line massive stars,  and 7\% for WR stars. Mauerhan et al. reached a detection rate of 95\%
for emission line  massive stars,  and up to 40\% for WR stars using  additional constraints
based on the  \Ks\ versus $J-$\Ks\  diagram, and X-ray emission.

An overview of near and mid-infrared colors of bright mass-losing stars in the  Milky Way is
missing,  and providing one is the purpose of  the present work.  In this work, we study
near-infrared and mid-infrared color properties of known  Galactic evolved stars-, and  define
salient color-criteria for selecting new  bonafide WR stars, RSGs, and AGBs. In Sect.\ 2, we
report on previous analyses of evolved  stars, which are based on GLIMPSE data. In Sect.\ 3,
we describe the samples of known  evolved stars, on which we base the new color-criteria. In
Sect.\ 4, we describe the assumed interstellar extinction ratios.  Color properties are
analyzed in Sect.\ 5, and discussed in Sect.\ 6.

\section{Sample selection} 
We collected  samples of known evolved massive stars (RSGs, WRs,
and LBVs)   with spectroscopically determined  spectral types. The lists of  AGBs were
compiled on the basis of  maser properties  and/or photometric  variability information
\citep[SiO masing stars, OH/IR stars,  Miras, and semiregular stars][]{habing96}.  

Throughout the text we  use the terms of late-type stars and early-type stars. These
definitions are only based on effective temperatures and not on  luminosity classes. AGBs and
RSGs are late-type stars, while LBVs and WRs are early-type stars.

For every object, we searched for counterparts in the 2MASS All-Sky Catalog of Point Sources 
\citep{cutri03}, in the third release of DENIS data available at CDS (catalog  B/denis) and 
in the GLIMPSE catalog \citep{spitzer09} using a search radius of 2\arcsec. The  II/293
(GLIMPSE) catalog available from CDS merges the three surveys GLIMPSE-I (v2.0), GLIMPSE-II
(v2.0), and GLIMPSE-3D; Catalog and Archive records are also merged. 

We searched for counterparts in the Version 2.3 of the MSX  Point Source Catalog (PSC) 
\citep{egan03} using a search radius of 5\arcsec. 

Only sources detected in at least one mid-infrared band were retained for further analysis.
Only 2MASS sources with good photometry were retained (i.e., with red flags  1, or 2, or 3, 
and quality flags A, B, C or D). Specific details on each sample are given in the following
sub-sections.

\subsection{Wolf Rayet stars}

Stars with initial masses larger than 40 \Msun\ enter the WR phase when  their H surface 
fractional mass abundance falls below 0.3 and their surface temperature is roughly above
20,000 K  \citep[e.g.][]{bressan84,figer97}. WRs  lose mass at  high rates  ($\sim$10$^{-5}$
M$_\odot$ $y^{-1}$) with their strong ionized winds, in which free-free emission is produced.
This emission can be detected as an infrared  excess long-ward of 2 \um\ \citep{cohen75}.  
Their spectral energy distribution can be described by a central star plus an infrared excess,
well described by a power law with a spectral index, $\alpha$, of $-0.6$, which corresponds to
the predicted index for a free-free emission generated in stellar winds \citep{wright75,
felli81}.

WC stars show circumstellar dust, that their  dust is likely due to  interactions in  multiple
systems \citep{waters10}.

We searched for GLIMPSE and MSX counterparts of a sample of 345 Galactic WR stars
\citep{vanderhucht01,messineo09, messineo11, mauerhan11,shara11}.  Coordinates by  Mauerhan et
al.\ and Messineo et al.\ are from 2MASS, while those from Shara et al. are from the The 
Naval Observatory Merged  Astrometric  Dataset  (NOMAD)  \citep{zacharias04}. The positions of
WR stars listed by \citet{vanderhucht01}   are  from  astrometric catalogs (e.g. Hipparcos),
but for a few stars less accurate  positions are given.   Initially, we searched for possible
associations within 10\arcsec; a few matches larger than  2\arcsec\ were individually checked,
and coordinates updated by using the spectroscopic catalog by \citet{skiff10}. The GLIMPSE and
2MASS astrometries agree within 1\arcsec. 

We retained the 120 GLIMPSE matches within 2\arcsec, and the 70 MSX matches within 5\arcsec.
The low number of GLIMPSE matches is mainly due to the survey  coverage (longitude
$l<65^\circ$  and latitude $b <1^\circ$). 195 out of 345 WRs are located in the range ($|l|<
65^\circ$ and $|b| <1^\circ$). 22 WRs stars have both MSX and GLIMPSE 8 \um\ measurements; the
mean value of their  ($[8.0]-[A]$) color is +0.22 mag with a standard deviation of 0.5 mag.
Discrepant measurements, more than 2.5 sigma, are  found for WR 125 and WR 93. Both sources
are colliding wind binaries of the same type (WC7 + O9) \citep{norci02}. For the dusty WR125
WR, variable mid-infrared flux  has been reported by \citet{williams94}.

\subsection{Luminous blue variables}

LBVs are rare massive stars  in transition  toward the Wolf-Rayet  phase
\citep[e.g.][]{clark05,conti95,martins07,nota95, smith04}.  They are characterized by cyclic
photometric variations with amplitudes of 1--2 mag and   yearly timescales, at nearly constant
bolometric luminosity. Giant eruptions with changes in  visual brightness by more than 3 mag
may also occur. LBVs have high mass-loss rates;  the archetypical LBV, $\eta$ Car, has a
mass-loss rate of $\sim10^{-4}$  \Msun\ yr$^{-1}$ \citep[e.g.][]{abraham05}.

LBVs may be surrounded by  extended dusty  nebulae, which show a variety of morphology and
composition.   Amorphous and cristalline silicate,  forsterite, as well as iron grains, have
been observed \citep{lamers01,waters10,umana10,peeters02}.   PAH emission are found to
dominate the $[8.0]-$band emission of two LBVs  \citep[e.g.][]{pasquali02,umana10,peeters02}.

We considered a list of 42  confirmed or candidate LBVs in the Milky Way \citep{clark05,
messineo11, gvaramadze10,  mauerhan10, messineo09}. We found 18 counterparts in the GLIMPSE
catalog and 25 in the  MSX  catalog. Five stars have 8 \um\ measurements in both the MSX and 
GLIMPSE surveys. However,     MSX measurements of WR102ka were discarded because of  high
crowding  in the galactic center region \citep{homeier03,barniske08}. For the remaining four
sources, the mean value of their  $[8.0]-[A]$ color  is +0.15 mag, with a standard deviation
of +0.19 mag.

\subsection{Red supergiant stars}

RSG stars are late-type  stars of initial masses between 8 and 40 \Msun, burning He in their 
core \citep[e.g.][]{levesque05}. RSGs have luminosities  log(L/\Lsun) from 4 to 5.3, often
show  irregular photometric variability, and  are surrounded by dusty circumstellar envelopes.
Their mass-loss rates range from  $\sim10^{-7}$ to $\sim10^{-4} $\Msun\ yr$^{-1}$
\citep[e.g.][]{josselin00,verhoelst09}.

Over 1000 stars are  listed as  RSGs in the spectroscopic catalog compiled by \citet{skiff10}.
However, several measurements are from the early 1900s, and need to be confirmed. A
significant number of RSG stars were recently discovered as members of   young massive
clusters  \citep[e.g.\ 2MASS and GLIMPSE][]{messineo09}. Five young massive stellar clusters
(RSGC1, RSGC2, RSGC3, RSGC4, and RSGC5),  extraordinarily  rich in RSGs (14, 26, $>8$, $>9$,
$7$), have been  located between 25\degr\ and 30\degr\ of  longitude, at a distance of about 6
kpc  probably at  the  near end-side of the Galactic Bar  \citep{davies07, figer06, clark09,
negueruela10, negueruela11}. We restricted our analysis to  a sample of  119 spectroscopically
confirmed RSGs  in clusters, because their association with a cluster confirms the luminosity
class \citep{skiff10,eggenberger02,mermilliod08,pierce00,mengel07,figer99,messineo11,
messineo10, davies09b,bernabei01, caron03,figer06, davies08,clark09, negueruela10,
negueruela11}. When comparing the distribution in colors of the 119 RSG stars in clusters with
all  RSGs listed by Skiff et al., however, no significant differences are seen. 

\citet{verhoelst09} studied the dust properties of O-rich RSGs envelopes and found a  dust
condensation sequence resembling that of AGBs \citep{kemper01, speck00, sylvester99}. Stars
with low mass-loss rates are rich of oxides and alumina, while high mass-loss stars show
olivine, and crystalline silicates \citep{waters10}. Metallic Fe or amorphous carbon could
also be present  \citep{verhoelst09, waters10}. \citet{buchanan09} report PAH emission at 6.2
\um\  in some LMC RSGs.  PAHs are not detected  in  mid-infrared spectra of Galactic RSG stars
obtained  with ISO-SWS \citep{sloan03}.

Among our sample of 119 RSGs, 69 matches were found with GLIMPSE counterparts and 79 with  
MSX  sources. 34 stars were detected by both the MSX and the GLIMPSE surveys.  The mean value
of their $[8.0]-[A]$ color is +0.11 mag with a standard deviation of 0.20 mag.

\subsection{Mass-losing AGB stars}

AGBs are stars with masses below 8 \Msun, with a degenerate core  consisting of C and O, and
two burning shells. The inner shell is burning He, while the  external one is burning H. The
presence of two different sources of nuclear energy determines an interplay between these two
shells, causing thermal pulses with inward and  outward motion of matter. Carbon  can be
brought onto the stellar surface, and   the AGB star is called a C-star, when the ratio of C
over O  is above 1.  The relative numbers of O and C-rich AGBs strongly depend on
metallicity.  We only consider here O-rich AGBs, since  AGBs in the  inner Galaxy are mostly
O-rich types \citep{sevenster99}.

Several different names are used to indicate specific sub-classes of AGBs, based on their
pulsation properties and/or circumstellar properties (pulsators, masing stars). A detailed
review on AGB stars is presented by \citet{habing03}. Miras are classically defined as AGBs
with  visual amplitudes larger than 2.5 mag;      their periods typically range  from  150 to
1500 days.  SRs are regular pulsators with visual amplitudes smaller that 2.5 mag; their
periods  range from 35 to 250 days. AGBs may  have circumstellar envelopes, where maser
emission  originates.  AGBs may be called SiO masing stars, or OH/IR stars, if they have SiO 
or OH maser emission.

Mass-loss in AGBs is caused  by stellar pulsation, and increases going from SR stars to Miras
and to  OH/IR stars, i.e.  with increasing periods.  It scales with  stellar luminosity
(\Mdot\ $\propto L^{2.7}$ in SRs)  and may range from $\sim10^{-8}$ to $\sim10^{-3} $ \Msun\
yr$^{-1}$ \citep[e.g.][]{alard01,habing03,ortiz02}. SR stars have typical mass-loss rates 
from $\sim10^{-8}$ to $\sim10^{-6} $ \Msun\ yr$^{-1}$, while Miras have  mass-loss rates from
$\sim10^{-7}$ to $\sim10^{-4} $ \Msun\ yr$^{-1}$. A median  mass-loss rate  of $3 \times
10^{-5}$ \Msun\ yr$^{-1}$ was estimated for a sample of OH/IR stars in the Galactic center
region  \citep[][and references therein]{habing03}.

In order to analyze the full range of near- and mid-infrared colors of AGBs, we collected 
samples of AGBs with known pulsational properties. About   $300$ SR stars were detected in 
the MACHO  and  ISOGAL surveys \citep{alcock99, alard01, schuller03}. We found 66 matches with
GLIMPSE  point sources and 17 with  MSX point sources. The 9 sources with available
($[8.0]-[A]$) colors yield a mean color of 0.45  mag with a standard deviation of 0.33 mag.

A sample of  409 large amplitude variables (LAV) was detected in the central $24\times24$
arcmin$^2$  of the Galaxy in $K$-band  by \citet{glass01}. LAVs  correspond to optical Miras,
with periods from 150d to 800d.  We found 291 counterparts to the 409 LAVs in the GLIMPSE
catalog and 95 in the MSX catalog. For the 71 LAVs with  measurements at $\sim8$ \um\ from
both MSX and GLIMPSE surveys,   the mean $[8.0]-[A]$ value is 0.24 mag with a standard
deviation of 0.8 mag. SiO maser was detected  in 77 of these LAVs  \citep[54 with a GLIMPSE
counterpart and 40 with an MSX counterpart,][]{imai02}.

The sample of SiO masing stars of \citet{messineo02} consists of 271 stars, which were color 
selected from the DENIS, ISOGAL, 2MASS and MSX catalogs.  Information on  variability  and
photometric properties suggests that the whole sample consists  of  Mira-like stars
\citep{messineo04}. This is a sample of  mostly AGB stars, but may likely includes a few RSGs
\citep{messineothesis}. Positions have an accuracy of 1\arcsec\ \citep{messineo04}.   For the
SiO masing stars, we found 177 GLIMPSE and 258 MSX counterparts. The 125 sources with both
8\um\ measurements have a mean  ($[8.0]-[A]$)  =0.34 mag and a standard deviation of 0.42 mag.

Along the text we will consider as an SiO maser sample the SiO masing stars by
\citet{messineo02} plus the 77 masing LAVs by  \citet{glass01,imai02}, while we will refer to
the remaining LAVs  by \citet{glass01} simply as  LAVs. Maser and color properties  are 
discussed in \citet{verheyen11}.

OH/IR stars are  large amplitude AGBs with periods generally longer than 600d (even exceeding
1500 days), and denser circumstellar envelopes than Miras or SRs.  The sample of 766 OH/IR
stars by \citet{sevenster02} consists of two  sets of maser stars detected with the VLA and
ATCA interferometers  ($|l| <45^\circ$ and  $|b|<3.5^\circ$). Positional uncertainties are
typically  1\arcsec. We  found   295 GLIMPSE counterparts and 610 MSX counterparts.  90 OH/IR
stars were detected  at 8\um\ by both MSX and GLIMPSE.  The mean ($[8.0]-[A]$)  is 0.35 mag
with a standard deviation of 0.52 mag. 

The different numbers of matches found in the GLIMPSE and MSX catalogs are due to   difference
in  their sensitivity and sky coverage. While MSX covers the whole Galactic plane to
$\pm5^\circ$ of latitude, the GLIMPSE survey is confined to a narrow latitude range
$\pm1^\circ$. OH/IR stars are  bright at 8\um; many were not included in the GLIMPSE catalog, 
because of saturation. The MSX detection limit in $A$-band  is 100 mJy ($[A]=6.9$ mag), while
the Spitzer/IRAC 8\um\ channel saturates at 444 mJy ($[A]=5.3$ mag), and has a 5 sigma
detection limit at 2mJy ([8]=11.2 mag).

\begin{figure*}[!]
\begin{centering}
\resizebox{0.8\hsize}{!}{\includegraphics[angle=0]{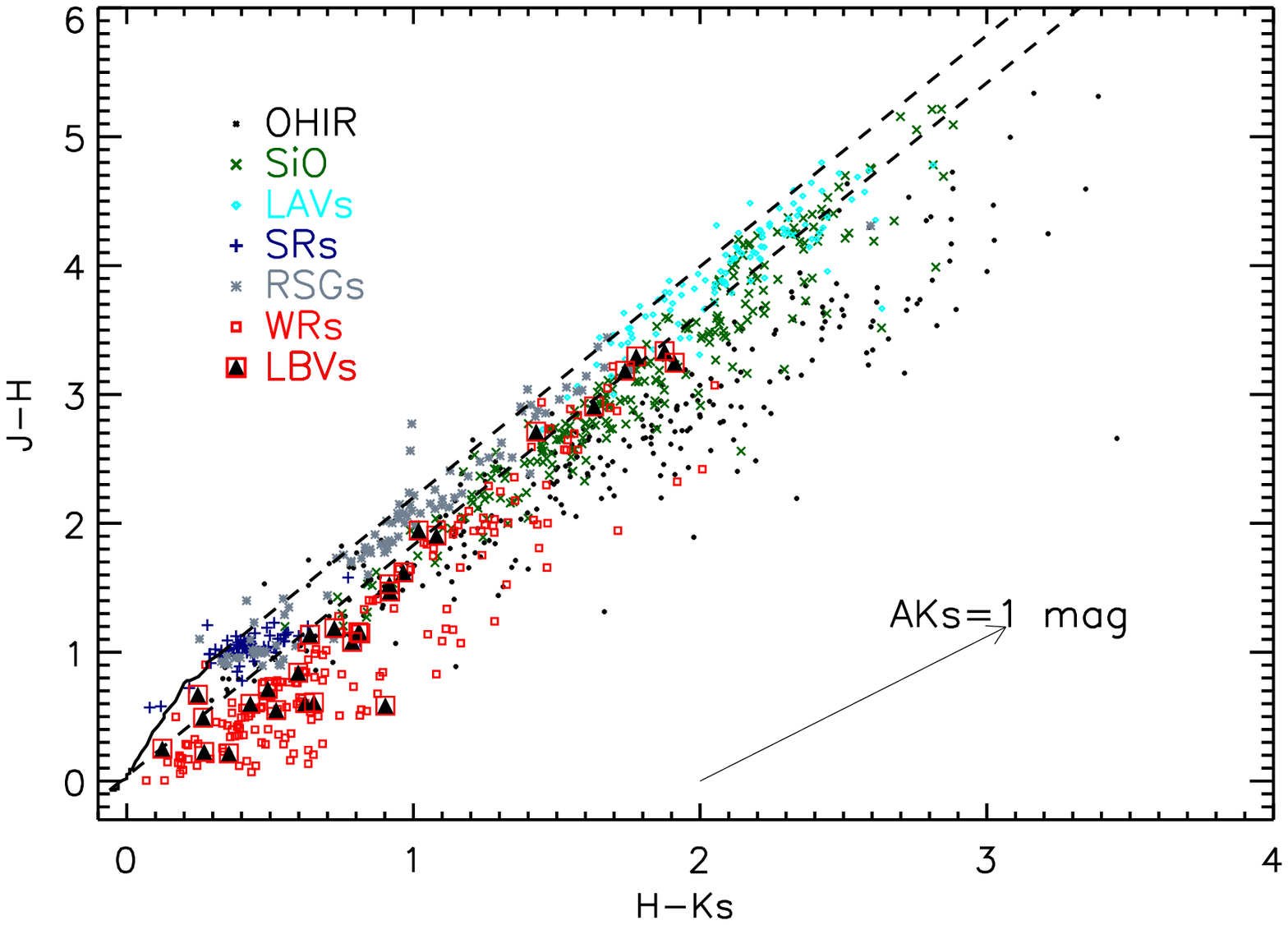}}
\end{centering}
\caption{\label{jhk} 2MASS $J-H$ vs. $H-$\Ks\ diagram of Galactic evolved stars.
This diagram does not allow for a photometric classification.
SR variables are taken from \citet{alard01}, 
large amplitude variables (LAVs) stars from \citet{glass01}, and
AGBs with SiO maser emission from \citet{messineo02}.
OH/IR stars are from \citet{sevenster02}. The selection of RSGs
is described into the text.
WR stars  are  from \citet{vanderhucht01,messineo09, messineo11}, LBVs 
are from \citet{clark05, messineo11, gvaramadze10,  mauerhan10, mauerhan11}. 
The continuous line near the bottom left corner is the locus of dwarf stars \citep{koornneef83},
two dashed lines mark the traces of an O9 (lower) and an M5 (upper) star with increasing 
interstellar extinction. A reddening vector for \Aks=1 mag is also shown \citep{messineo05}.
Typical  errors are within  0.05 mag in both axis.}
\end{figure*}

\begin{figure*}[!]
\begin{centering}
\resizebox{0.8\hsize}{!}{\includegraphics[angle=0]{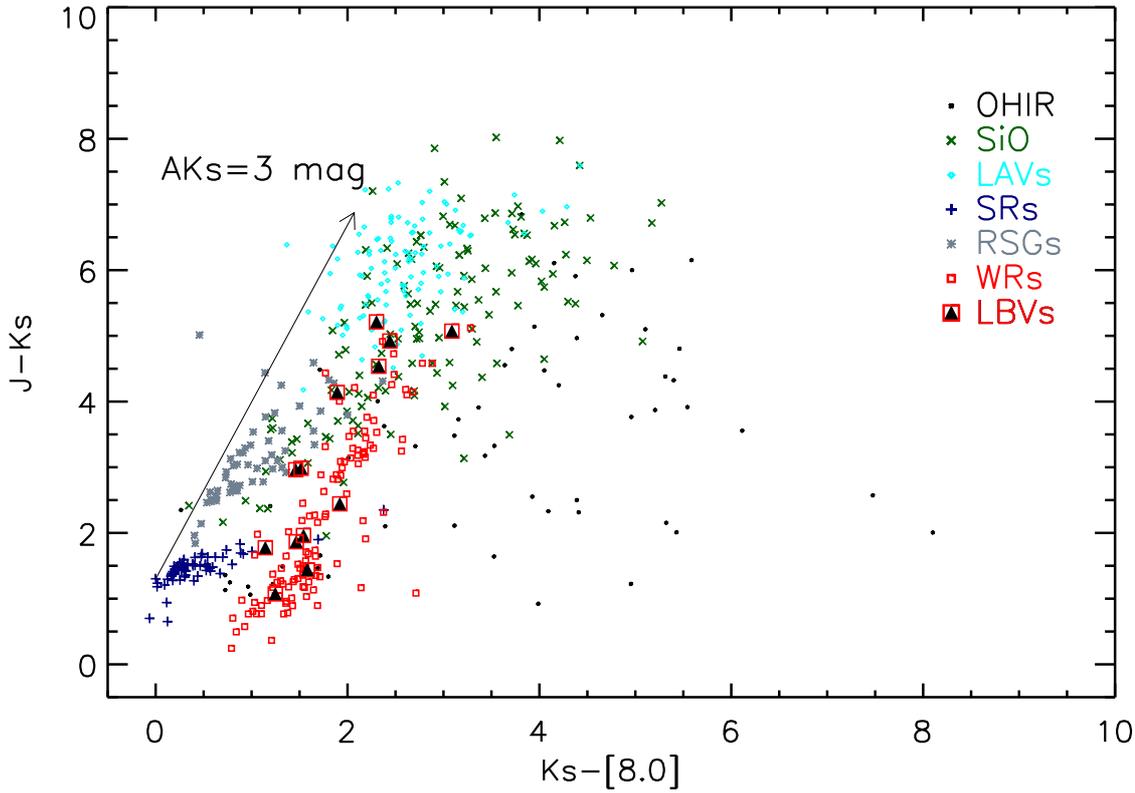}}
\end{centering}
\caption{\label{lucy2} 2MASS/GLIMPSE \Ks$-[8.0]$ versus $J-$\Ks\ diagram.
Symbols are as in Fig.\ \ref{jhk}. The arrow represents the reddening vector
following the extinction ratios given by \citet{messineo05} and \citet{indebetouw05}.
Typical  errors are within  0.07 mag in both axis.}
\end{figure*}

\begin{figure*}[!]
\begin{centering}
\resizebox{0.8\hsize}{!}{\includegraphics[angle=0]{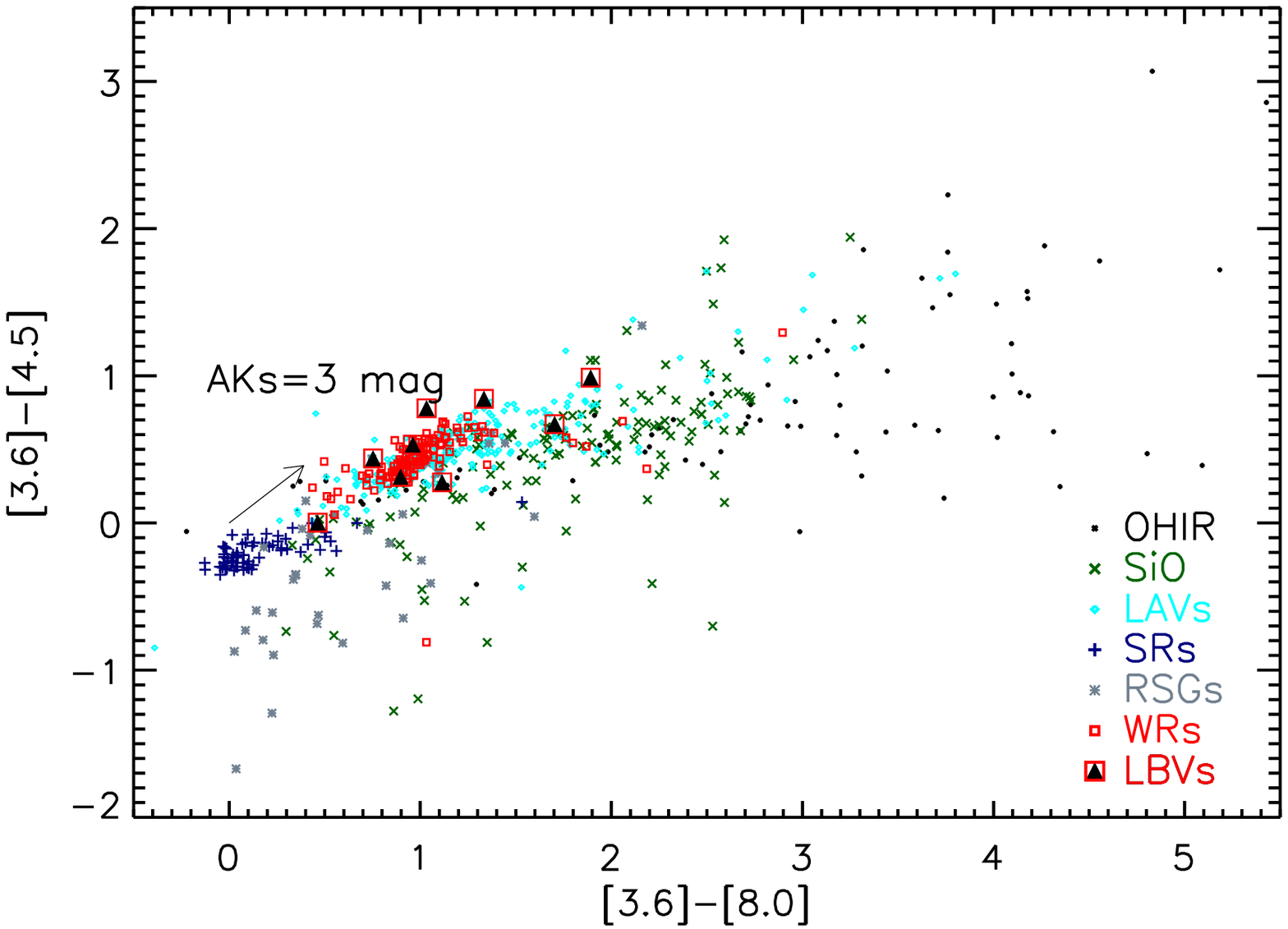}}
\end{centering}
\caption{\label{lucy1} GLIMPSE $[3.6]-[8.0]$ versus $[3.6]-[4.5]$ diagram.
Symbols are as in Fig.\ \ref{jhk}.
The arrow indicates the direction of the reddening vector following the extinction ratios
derived by \citet{indebetouw05}.
Typical  errors are within  0.08 mag in both axis.}
\end{figure*}

\section{Interstellar extinction}

We adopt a near-infrared extinction power law ( \Ak\ $ \propto \lambda ^{-\alpha}$)
\citep{messineo05,rieke85,indebetouw05,nishiyama06, stead09,fritz11}, with  an index of
$\alpha=-1.9$ \citep{messineo05}.  This index provides consistent measurements of extinction
in the  \Ks\ versus ($J-$\Ks) and \Ks\  versus ($H-$\Ks)  diagram, also in fields with very
high  extinction \Aks $>2$ mag.  Furthermore, this index  is consistent with that measured
with hydrogen recombination lines by \citet{landini84}. A recent historical review of  index
measurements  is presented by \citet{fritz11}.

For the GLIMPSE measurements, we used the color excess ratios $E_{\lambda-Ks}/E_{J-Ks}$ 
calculated  by  \citet{indebetouw05}, and we re-calculated the  extinction ratios using Eqn.\
1 in \citet{indebetouw05}, and   $A_J/A_{Ks}$ from \citet{messineo05}. The assumed and
resulting quantities are listed in Table \ref{AlambdaAk}. They are lower than those first
reported by \citet{indebetouw05}.  Lower values are also  measured toward the Galactic center 
\citep{fritz11}.

The use of  slightly different extinction ratios does not affect our analysis,  because the 
average excess colors are much larger than the uncertainties due to interstellar extinction.

\begin{table*}

\caption{\label{AlambdaAk}  Infrared extinction ratios (A${_[\lambda]}$/\Aks) for the 2MASS  
and  SPITZER/IRAC filters.}

\begin{tabular}{lllllll}
$A_J/A_{Ks}$  & $A_H/A_{Ks}$    & $A_{[3.6]}/A_{Ks}$      & $A_{[4.5]}/A_{Ks}$     & $A_{[5.8]}/A_{Ks}$      & $A_{[8.0]}/A_{Ks}$ & Reference \\
2.86          & 1.66            &               &              &               &               & \citet{messineo05}   \\
2.5           & 1.56            &               &              &               &               & \citet{rieke85} \\
$2.5\pm0.2$   & $1.55\pm0.1$    & $0.56\pm0.06$ & $0.43\pm0.08$& $0.43\pm0.10$ & $0.43\pm0.10$ & \citet{indebetouw05}\\
2.86          & 1.66            &0.44           &0.31          & 0.29          &  0.31         & \citet{indebetouw05} and \citet{messineo05} \\
\end{tabular}
\end{table*}

\section{Color-color diagrams}

In order to identify photometric criteria that might separate early-type stars from late-type 
stars and identify specific types of stars (e.g.\ WRs and RSGs), we analyzed the distribution
of  evolved stars in several color-color diagrams. 

\subsection{$H-$\Ks\  versus $J-H$ diagram}

Because of their different temperatures, "normal" late-type  stars and early-type stars  form
two parallel sequences with increasing interstellar extinction in the $H-$\Ks\  versus $J-H$
diagram.  This diagnostic tool    is suited to only classify the bulk of a stellar population;
it relies on the assumption that mass-losing objects are rare. Mass-loss changes the stellar
energy distribution.

Dust grains in the circumstellar envelopes of late-type stars absorb the stellar light
 and re-emit at longer wavelengths \citep[e.g.][]{groenewegen93}.
In the solar neighborhood,  AGBs with low mass-loss rates $<10^{-7}$ \Msun\ yr$^{-1}$ have 
de-reddened  $J-$\Ks\ colors, $(J-K)_{o}$, between 1.2 and 1.6 mag, while AGB with  mass-loss  from 
$10^{-6} - 10^{-4}$ have $(J-K)_{o}$  from 2 to 6.5 mag  \citep[][and references therein]{messineo05}.
AGBs can have up to 40 mag of visual extinction.
Figure  \ref{jhk} shows  the $H-$\Ks\ versus $J-H$ diagram of our samples  
of evolved stars. Two continuous lines are plotted, showing
the loci of a "naked" M5 star (left/higher dashed line) and of an O9 star (right/lower line) \citep{koornneef83} with 
varying interstellar extinction, as well as the reddening vector \citep{messineo05}.
It is not possible to classify stars with this diagram alone, e.g.\ all Miras would be selected as OB stars.
RSGs, SRs, and Galactic center LAV stars 
appear located between these two lines. SiO and OH/IR stars  show even redder $H-$\Ks\ colors,
which are  due to continuum absorption in $H$-band  by water
\citep[e.g., Fig. 6 and 7 in ][]{messineo05, blum03,frogel87, comeron04}.  

WRs seem to define a  sequence in Fig. \ref{jhk},  
which lies below the sequence of "normal" OB stars ($H-$\Ks\ from  0.0 to 1.5 mag and $J-$\Ks\ 
from 0.0 to 2.0 mag). The distribution of WRs overlaps  that of  evolved late-type stars 
(OH/IR and SiO masing stars). Different physical mechanisms can generate  similar $H-$\Ks\ 
and $J-H$ colors; LBVs  have $J-$\Ks\ colors from $\sim 1.0$ to $\sim 5.0$ mag, 
and $H-$\Ks\ from $\sim 0.2$ mag to $\sim 2.0$ mag. A sample 
of five LBVs (AFGL2298, Pistol Star, WR102ka, LBV1806$-$20, G0.120$-$0.048, FMM362)
have redder $J-$\Ks, which are consistent with distant sources  at large 
interstellar extinction 
\citep[\Aks=3.15, 2.99, 2.78, 3.0, 3.26, 3.42 mag, respectively][]{clark03,mauerhan10,barniske08,bibby08}.

Since circumstellar absorption moves the stars on the diagram almost along the 
direction of the interstellar reddening vector, it is impossible to distinguish 
between  circumstellar and interstellar reddening with this diagram. 
An estimate  of total (interstellar plus circumstellar) extinction can be obtained 
from the reddening in $J-$\Ks, \Ak$_{J-Ks}= E((J-$\Ks$))*0.54$ \citep{messineo05}.

\subsection{\Ks$-[8.0]$ versus $J-$\Ks\ diagram}
A useful diagnostic for mass-loss is a  \Ks$-X$ versus $J-$\Ks\ diagram,
where $X$ is a mid-infrared magnitude, e.g. the 15 \um\ measurements made by
ISOGAL, or the 8 \um\ measurements made by MSX and GLIMPSE \citep[e.g.][]{messineo05,vanloon03}.
In Fig.\ \ref{lucy2}, we show a  \Ks$-[8.0]$ versus $J-$\Ks\ diagram.
The reddening vector is determined using the near-infrared extinction 
law by \citet{messineo05}, and the color excess by \citet{indebetouw05} (see Sect.\ 4).
A separation between interstellar and circumstellar extinction is detected in this diagram,
since increasing circumstellar reddening do not move the star along the direction of the 
interstellar reddening vector. Deviations from the reddening vector provide evidence for   
mass-loss  \citep[e.g.][ and references therein]{messineo05}.

SRs are almost located on the reddening vector, while OH/IR stars
show large  \Ks$-[8.0]$ color excesses (up to 8.0 mag). 
There is a large overlap   between the colors of AGBs (OH/IR stars, SiO masing stars, 
and Miras) and  RSGs stars.   

WRs  follow a separate sequence in the \Ks$-[8.0]$ versus $J-$\Ks\ diagram \citep{hadfield07}.
This WR sequence suffers  contamination from late-type stars.

\subsection{ $[3.6]-[8.0]$ versus $[3.6]-[4.5]$  diagram}
In Fig.\ \ref{lucy1}, we show a  $[3.6]-[8.0]$ versus 
$[3.6]-[4.5]$ diagram of evolved stars.
The reddening vector from  \citet{indebetouw05} is also marked.
Evolved stars do not follow the interstellar reddening vector, but, for a given
$[3.6]-[4.5]$ color, they display  redder $[3.6]-[8.0]$ colors. 
We expect interstellar extinction in the IRAC bands below 0.39 mag 
even in the central regions of the Milky Way (\Ak $\sim$ 3 mag) \citep{indebetouw05}. 
Since the ranges of colors seen in Fig. \ref{lucy1} are much  larger than this value,
we can extract indication on intrinsic colors for each stellar type.

In Fig.\ \ref{lucy1}, AGBs (SRs,  Miras, and masing stars) form a sequence of 
increasing $[3.6]-[8.0]$ colors,  from 0.0  (SRs) to 5.0 mag (OH/IR stars).
This is likely due to a combination of increasing mass-loss rates and 
water absorption. 
SR and LAVs stars  display increasing $[3.6]-[4.5]$ colors, with increasing
$[3.6]-[8.0]$ colors, likely due to increasing strengths of  water absorption at $\sim 3$\um. 
SiO masing stars  display peculiar $[3.6]-[4.5]$ colors, bluer than those of SRs and 
LAVs (without SiO masers). A possible cause is a difference in the water strengths 
and possible SiO absorption  at $\sim 4$ \um\ and CO$_2$ absorption at $\sim 4.3$ \um
\citep{cami02}.
The distribution of RSGs resembles that of masing AGB stars in this plane.
Some RSGs  have the  bluest $[3.6]-[4.5]$ colors than any other stellar type, as already 
noted in the LMC, likely due to CO \citep[e.g.][]{yang11}. 

Few ISO-SWS spectra of late-type stars \citep{sloan03} are shown in Appendix.

The bulk of WR stars have $[3.6]-[8.0]$  from 0.5 to 1.5  mag and $[3.6]-[4.5]$ from
0.0 to 0.7 mag. The colors of early type stars, WRs and LBVs, overlap with those 
of late type stars.

\section{The $Q1$ and $Q2$ parameters}

\begin{table}[!]
\caption{ \label{tableq1}
 $Q1$ values for normal stars (without mass-loss).}
\begin{tabular}{rrrrrrr}
\hline\\
Sp. type & I & $Q1$ & III & $Q1$& V& $Q1$\\
\hline\\
  O9&I&$-$0.02&III&     &V&$-$0.1\\
  A0&I&   0.07&III&     &V&0.01\\   
  F0&I&   0.13&III&     &V&0.13\\   
  G3&I&   0.31&III&0.36 &V&0.23\\   
  K0&I&   0.36&III&0.41 &V&0.29\\   
  M5&I&   0.72&III&0.68 &V&0.38\\  	
  \hline\\
\end{tabular}
\end{table}

\subsection{The $Q1$ parameter}

\begin{figure*}[!]
\begin{centering}
\resizebox{0.8\hsize}{!}{\includegraphics[angle=0]{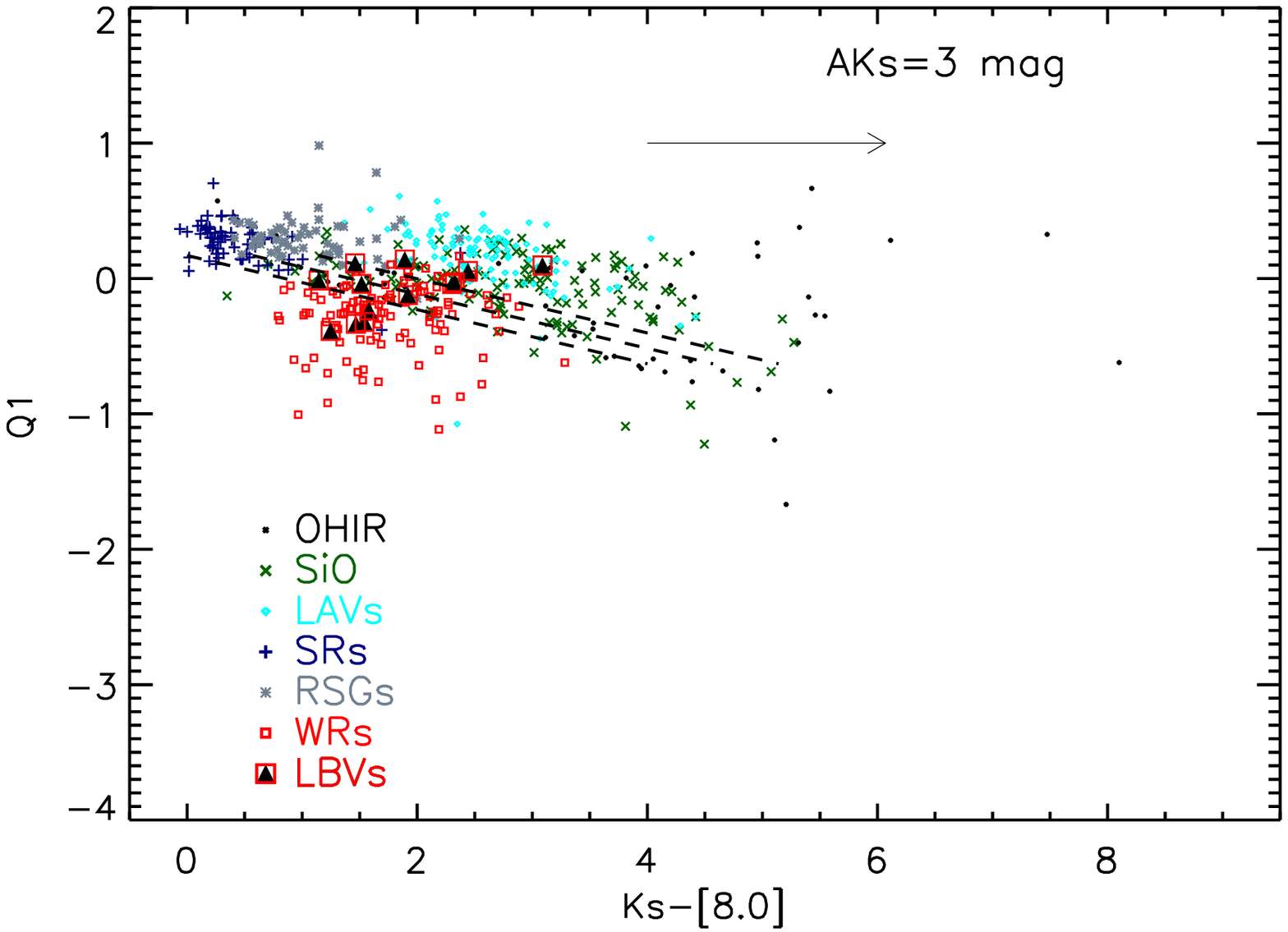}}
\end{centering}
\begin{centering}
\resizebox{0.8\hsize}{!}{\includegraphics[angle=0]{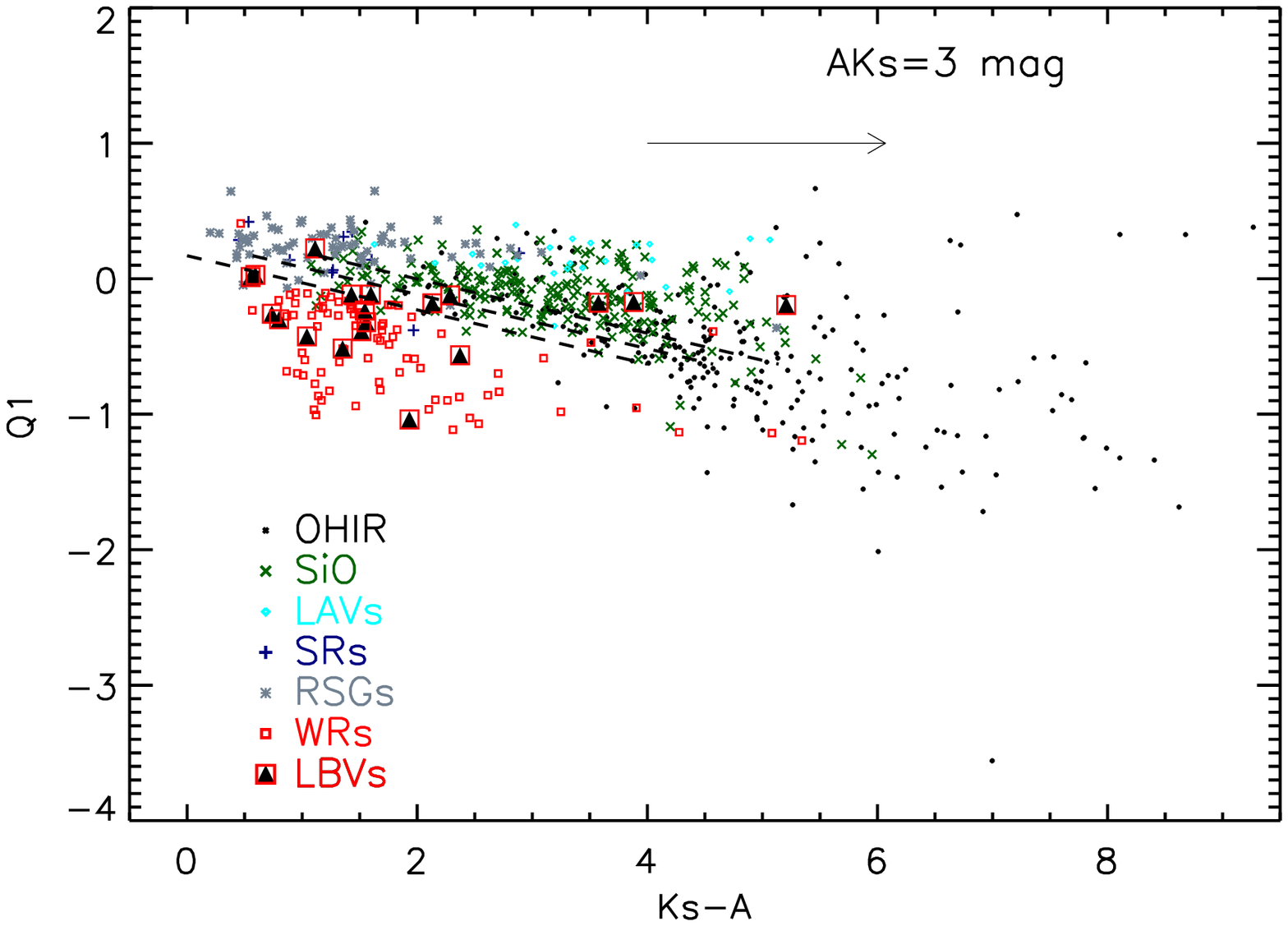}}
\end{centering}
\caption{\label{q} $Q1$ parameter versus \Ks$-[8.0]$. Symbols are as in Fig. \ref{jhk}.
In the top panel the GLIMPSE 8 \um\ measurements are plotted, while
in the bottom panel the MSX 8 \um\  measurements ($A$-band) are plotted.
The arrow represents the reddening vector
following the extinction ratios given by \citet{messineo05} and \citet{indebetouw05}.
The dashed lines  are a linear fit to the $Q1$ versus the dereddened (\Ks$-[8.0]$)
values of SiO masing stars \citep{messineothesis} for \Ak=0,1,2 mag (from bottom to top).
Typical  errors are  0.05 mag in the x-axis and 0.14 mag in the y-axis.}  
\end{figure*}

The $Q1$ parameter, $Q1=(J-H)-1.8 \times (H-$\Ks), is a measure of the deviation from the
reddening vector in the $H-$\Ks\ versus $J-$\Ks\ plane, following the infrared extinction 
law by \citet{messineo05}. 
This quantity does not depend on interstellar extinction.
The $Q1$ parameter was first introduced  to photometrically select 
counterparts of high-mass X-ray binaries \citep{negueruela07}; it has been successfully 
used for   searching clusters of RSGs \citep[e.g.][]{figer06, davies07, clark09,negueruela10,negueruela11}.
Since stars with different spectral types have different intrinsic colors, they will also
have a different value of $Q1$, as shown in Table \ref{tableq1}  with   intrinsic colors by \citet{koornneef83}. 
$Q1$ is typically around 
zero for "normal" OB stars, while it is around 0.4 mag for K-giant stars \citep{koornneef83}.

An infrared excess due to free-free emission in the winds of hot stars (e.g. WRs), or to dusty
circumstellar envelopes (e.g. AGBs or RSGs), can modify the $Q1$ parameter of stars. 
Mass-losing late-type stars may have negative $Q1$ values 
(similar to those of early type stars). The $Q1$ parameter alone does not allow distinguishing 
between early and late-type stars.

\begin{table}[!]
\caption{\label{q1q2table}  Typical $Q1$ and $Q2$  ranges for the considered sample of stars.
}
\begin{tabular}{rrrrr}
Type     & $Q1$ (min) & $Q1$ (max)  &$Q2$ (min) & $Q2$ (max)   \\
\hline
    RSG &  $-$0.06&  0.44 &   $-$3.38&  1.12 \\
     WR &  $-$1.19&  0.11 &   $-$5.02& $-$1.02 \\
    LBV &  $-$0.34&  0.06 &   $-$2.23& $-$0.73 \\
  OH/IR &  $-$1.56&  0.44 &  $-$12.99& $-$0.49 \\
    SIO &  $-$0.60&  0.30 &   $-$7.31&  0.69 \\
    LAV &  $-$0.17&  0.53 &   $-$3.01&  1.49 \\
     SR &     0.02&  0.52 &   $-$1.04&  1.46 \\
\hline
\end{tabular}
\end{table}

In Fig.\  \ref{q}, the $Q1$ parameter is plotted against the \Ks$-[8.0]$ color.
Two panels are shown to separately plot the $Q1$ versus the \Ks$-[8.0]$  colors from
the 2MASS and GLIMPSE catalogs, and the $Q1$ versus \Ks$-A$ colors from the 2MASS and MSX catalogs. 
The numbers of stars per type  are different in the two panels. Most of the OH/IR stars 
are detected by MSX, but are saturated in GLIMPSE. 
Galactic center SRs  could only be detected  with the more sensitive GLIMPSE survey.
Typically, $Q1$ is less than 0.1 mag for windy early-type stars, and  ranges from  $-0.1$ to $0.4$ mag 
for the bulk of known RSG stars. AGB SRs and AGB LAVs have  similar ranges of $Q1$ 
values, which are also similar to those of  RSGs. AGBs with SiO or OH maser emission  
have broader $Q1$ ranges, OH/IR stars have values down to $\sim -1.6$ mag (see Table \ref{q1q2table}).
Late-type stars move to redder \Ks$-[8.0]$ with decreasing $Q1$ values.
Most of the WRs  have negative $Q1$ values. LBVs have $Q1$ values range from 0.1 to $-0.35$ mag.

In Fig.\ \ref{q} the $Q1$ values are interstellar extinction independent, while the \Ks$-[8.0]$
and \Ks$-A$ values are  affected by interstellar reddening (E(\Ks$-[8.0]$)),  
and, therefore, interstellar reddening can only move a data point  on a horizontal line.
We detected  a diagonal cut-off in the distribution of $Q1$ versus \Ks$-[8.0]$ values of 
late-type stars in Fig.\ \ref{q}.  This is unrelated to interstellar extinction, and 
arises from the intrinsic stellar energy 
distribution (SED) of late type stars. The stellar energy distribution significantly changes 
with increasing mass-loss also in the $J$, $H$, and \Ks\ bands. Bolometric corrections 
in \Ks-band (\BCKs)  and in $A$-band (\BCA) as a function of \Ks$-A$  were calculated for a 
sample of SiO masing stars in the inner Galaxy, and compared with those of  AGBs in the solar 
neighbours \citep[Chapter5, Appendix A2 of ][]{messineothesis, whitelock00,
whitelock94, whitelock03, olivier01}. 
\BCKs\  decreases with increasing \Ks$-A$, and  for  \Ks$-A$=6 mag \BCKs$=0.8$ mag, i.e.\  
$\sim 2.3$ mag smaller than  values  for "naked"  M-type stars 
($2.9-3.1$ mag) \citep[e.g.][]{blum03,frogel87}.

 We obtained the following relations between the $Q1$ and \Ks$-[8.0]$
values of SiO masing Mira-like stars \citep[][]{messineothesis}: \\

$Q1= 0.50(\pm0.17) -0.22(\pm0.05)\times BC_A$

$Q1= 0.34(\pm0.05) -0.20(\pm 0.04) \times(K_S-A)_o$.
\noindent
where \Ks$-A$ ranges from $\sim 0$ to $\sim 4$ mag. 

These relations prove that changes in the SEDs cause variations in the 
$Q1$ values, and generate the cut-off seen in  Fig.\ \ref{q}. 
Reddening in \Ks$-A$, due to interstellar extinction, and stellar pulsation smooth 
such a cut-off by smearing it over \Ks$-A= \sim 1.5$ mag. 
For an \Aks\ of 1, 2, and 3 mag, E(\Ks$-[8.0])=\sim 0.31$, $\sim 0.62$, and 
$\sim 0.93$ mag, respectively.
This relation nicely explains the diagonal cut-off observed in Fig.\ \ref{q}. 
OH/IR stars with  much redder colors (\Ks$-A > 4.0$ mag) may fall below the 
relation found for SiO masing stars.

This relation between  $Q1$ values and \Ks$-[8.0]$ colors 
allows a  separation of  mass-losing late-type stars from hot
windy stars  (e.g. WR stars). A full revision of bolometric corrections
in several near and mid-infrared bands will be the topic of a forthcoming paper.
 
A few peculiar WRs  fall above these lines;  WR 125 has a $Q1$ value 0.8 mag above 
this line when using its MSX 8 \um\ measurement. This source is known to be dusty 
and variable \citep{williams94}.  LBVs cannot be identified in this diagram.
Their $(K_S-A)_o$ colors expand over the region of WR stars, as well as over that 
of late-type stars.
 
\subsection{The $Q2$ parameter}

\begin{figure*}[!]
\begin{centering}
\resizebox{0.8\hsize}{!}{\includegraphics[angle=0]{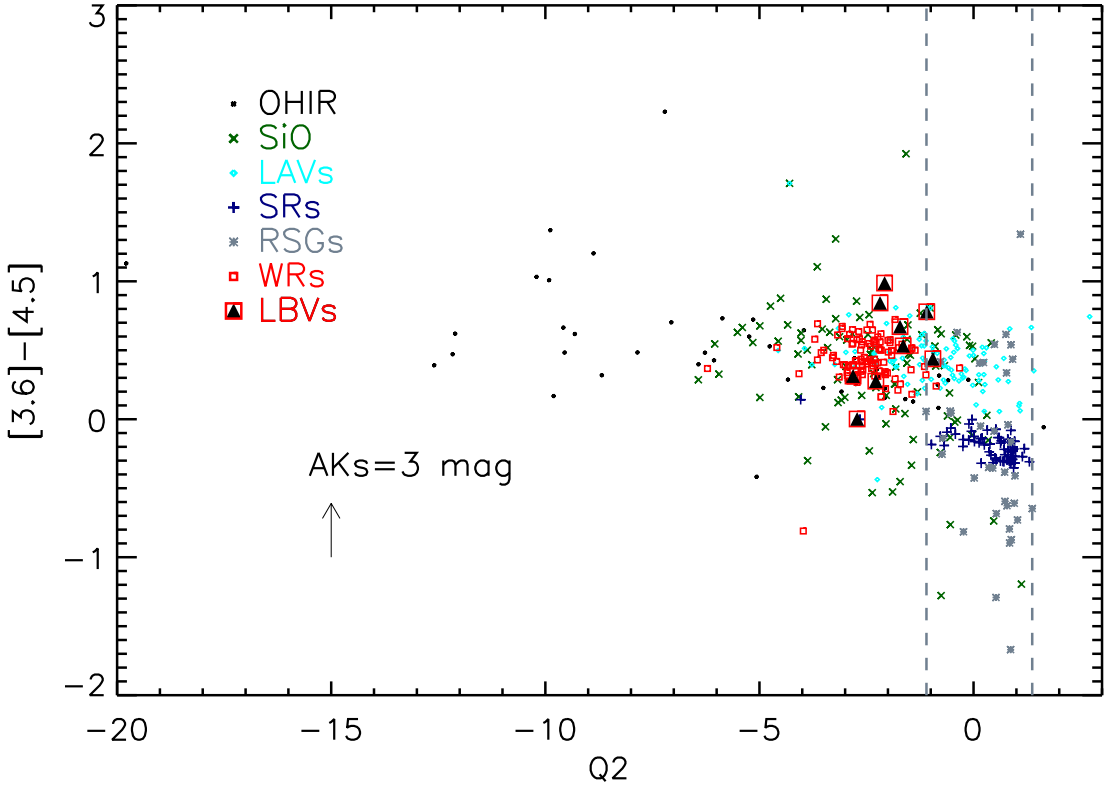}}
\end{centering}
\caption{ \label{q2} $Q2$ parameter versus the GLIMPSE $[3.6]-[4.5]$ color. 
Symbols are as in Fig.\ \ref{jhk}.
The arrow represents the reddening vector
following the extinction ratios given by  \citet{indebetouw05}.
Dashed vertical lines at $Q2$=$-1.1$ mag and $Q2$=$1.37$ mag indicate 
the region populated by RSG stars. Stars in this range and with $[3.6]-[4.5] <-0.4$ 
mag are highly probable RSG stars. 
Typical  errors are  0.13 mag in the x-axis and 0.08 mag in the y-axis.}
\end{figure*}

We define a new  parameter $Q2=(J-$\Ks$)-2.69\times$(\Ks$-[8.0])$, which is a measure of 
the deviation from the reddening vector in the
$(J-$\Ks) versus \Ks$-[8.0]$ plane, following the infrared extinction law of \citet{messineo05}
and mid-infrared color excess ratios by \citet{indebetouw05}.  $Q2$ can be thought of  as a 
measure of an excess that is  only due to a circumstellar shell (E(\Ks$-[8.0])_{shell}$), 
and is independent of interstellar extinction: $Q2 \propto$ (E(\Ks$-[8.0])_{shell}$).

We prefer to use the excess in the $(J-$\Ks) versus (\Ks$-[8.0])$, rather than  in the 
$(H-$\Ks) versus (\Ks$-[8.0])$ plane,  because AGBs may have strong absorption in $H$-band 
by  gaseous water  \citep[see Fig. \ref{jhk}, or][]{messineo05, blum03, frogel87}.

The $Q2$ parameter can be used to statistically select stars with infrared excess. 
Such excess can be due to dust emission  (late type stars), or  
free-free emission in shocked winds (WRs);
for LBVs it is generally due to free-free emission plus dust emission.

In Fig.\  \ref{q2}, the $Q2$ parameter is plotted against the $[3.6]-[4.5]$ color.
$Q2$ is a  parameter  independent of interstellar extinction, 
interstellar reddening in the $[3.6]-[4.5]$ color are small because 
the extinction ratios $A_{[3.6]}$/\Aks=0.44 and $A_{[4.5]}$/\Aks=0.31 are similar \citep{indebetouw05}.
The reddening in $[3.6]-[4.5]$ is 0.13, 0.26, and 0.39 mag, for \Aks=1, 2, 3 mag,  respectively.

The samples of masing stars (which are mostly Mira-like stars) show typical $Q2$ values 
from $\sim  -13$ mag to $\sim 0.7$ mag, but mostly smaller than $-1.0$ mag. SR stars and 
RSGs have mostly $Q2$  between 1.5 and $-1.1$ mag. By combining color and magnitude 
information it is possible to select  bonafide candidate RSG stars, since about 70\% 
of the stars in the range of $Q2$ from  $-1.1$  and 1.5 mag, and $[3.6]-[4.5] < -0.4$ mag, are RSGs.
WR stars have $Q2$ values  from  $-$5.0 mag to $-$1.0. LBVs have a narrow range of $Q2$ 
values from  $-$2.2 to  $-$0.7 mag.

\subsection{$Q1$ versus $Q2$}
\begin{figure*}[!]
\begin{centering}
\resizebox{0.8\hsize}{!}{\includegraphics[angle=0]{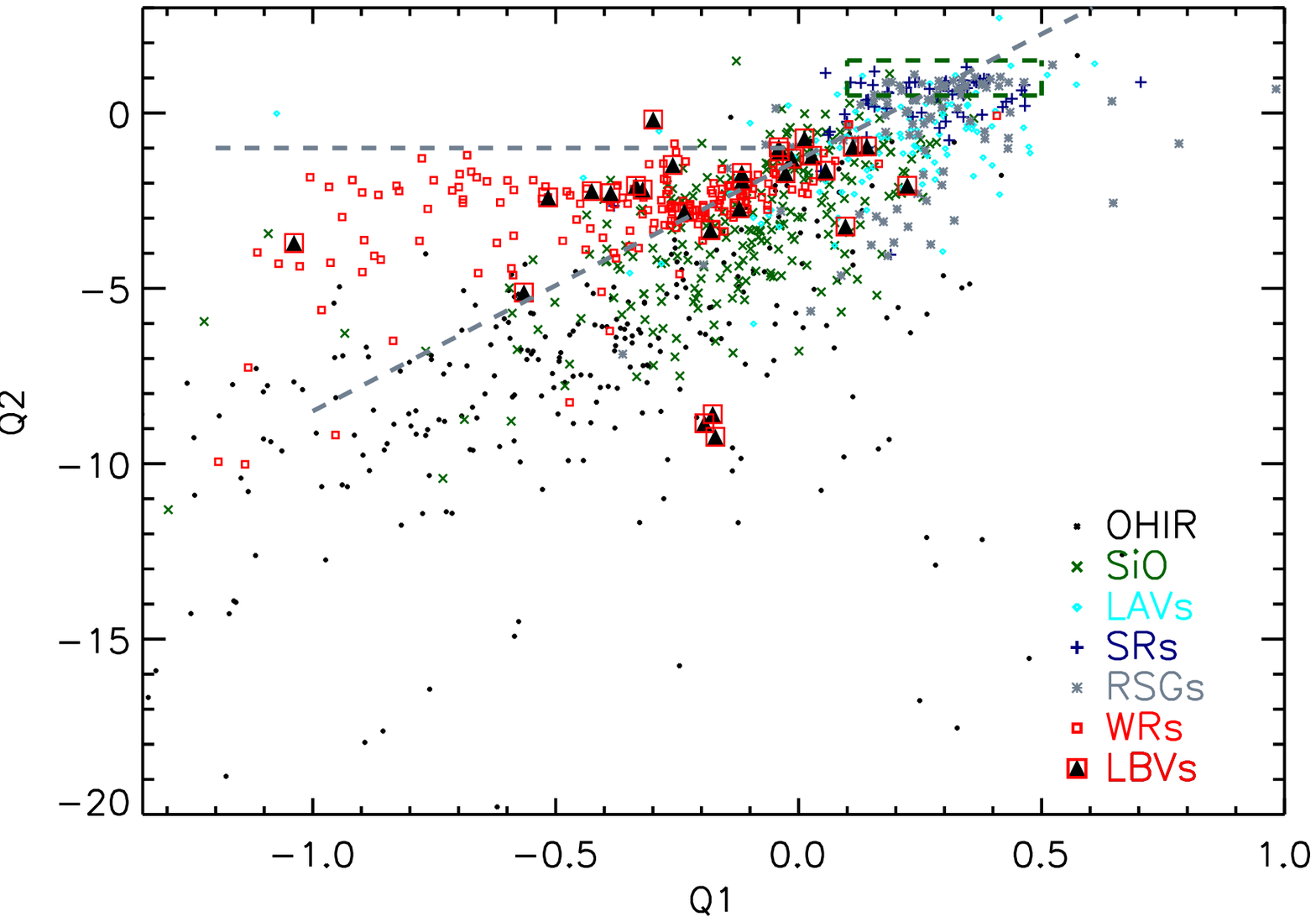}}
\end{centering}
\caption{\label{figq1q2} $Q1$ versus $Q2$ values, which are values independent of interstellar extinction.
Plotted symbols are as in Fig.\ \ref{jhk}. 
The dashed line ($Q2 = 11.25 \times Q1 -2.38 $ mag) serves as a first approximation to separate early and late-type
stars. The dashed box with  $0.1 < Q1 < 0.5$ mag and  $0.5 < Q2 < 1.5$ mag indicates the
selection of candidate RSGs (see text RSG2). WR stars populate the region enclosed by the two
lines $Q2 > 11.25 \times Q1 -2.38 $ mag and $Q2 < -1.0 $ mag. }
\end{figure*}

 In Fig.\ \ref{figq1q2} we plot the $Q1$ versus $Q2$ values  for the samples of  known evolved
stars described in Sect.\ 3.  We estimated the $Q1$ and $Q2$ ranges of Table \ref{q1q2table}
by making histograms of their  values and retaining only bins with at least 4 elements; 
the enclosed fractions are generally above 90\%.

Early type stars (with the exception of LBVs) and late type stars are distributed differently in the 
$Q1$ versus $Q2$ diagram. Early-type stars are mostly located above the relation  
$Q2=11.25*Q1-2.38$ mag, while mass-losing late-types (with the exception of SRs) are mostly located 
below it. However, a clearer separation between early and late-type stars appears in the $Q1$ vs $K-A$ 
diagram (see Fig.\ \ref{q}), where
early and late-type mass-losing stars can be separated with a 20\% error (see Table \ref{result}).
Specific searches can be optimized by adopting more  stringent constraints on $Q1$ and $Q2$, 
and by using  luminosity information. 

AGBs form a sequence of  decreasing $Q1$ and $Q2$ values going from SRs to OH/IR stars.
The bulk of SRs have $Q1$ from 0.0 to +0.5   mag, and $Q2$ from $-1.0$ to $1.5$  mag. 
LAV stars have similar $Q1$ values, but  their $Q2$ values  have a larger range (from $-3.0$ to $1.5$ mag).
Masing stars have a slightly larger range of $Q1$, from $-1.6$ mag to 0.4 mag, and generally negative $Q2$,
down to $-13.0$ mag. 

The $Q1$ and $Q2$ values of RSGs are similar to those of  LAV stars. Their $Q1$ values range from $-0.0$ 
to $0.4$ mag, and $Q2$ are mostly from  $-3.4$ to $1.1$ mag.
A large fraction (56 \%) of the RSGs appear concentrated in the region  $0.1< Q1 < 0.5$ mag and $-1.1 < Q2 <  1.5$ mag,
and 26 \% in the narrower region $0.1< Q1 < 0.5$ mag and $0.5 < Q2 <  1.5$ mag.
The fraction of LAVs and masing stars falling in the latter region is  below 2\%.
SR stars cannot be distinguished from RSGs on the basis of colors only 
(49\% of SRs  fall into the narrow region). Since SRs are intrinsically much fainter, the degeneracy 
can be eliminated with luminosity information.

WR stars have distinct $Q1$ and $Q2$ values; typical $Q1$ values are smaller than $\sim 0.1$ mag
and $Q2$ values smaller than $-1.0$ mag.
LBVs have typical $Q1$ values from $-0.3$ to 0.0 mag, and $Q2$ values from $-2.2$ to $-0.7$ mag.
A distinct group of three LBVs (Wra751/IRAS11065-6026, AFGL2298, and	
HD168625) appears  at $Q1 \sim -0.2$ mag and $Q2 \sim -9.0$ mag.
HD 168625 and AFGL2298 have large $Q2$ values because of their strong PAH emission lines 
\citep{pasquali02,umana10,peeters02}.
This suggests that the spectrum of Wra751/IRAS11065-6026 is  also dominated by strong PAHs.

\section{Suggested selection criteria}
\label{types}

There is overlap among the various  samples  in the color-color diagrams shown 
in Figs. \ref{jhk}, \ref{lucy2}, and \ref{lucy1}. It is not possible 
to uniquely define a  photometric classification scheme that is only based on these diagrams.
Furthermore, each color depends on interstellar extinction.  
 
We propose a  new classification  based on two extinction free  parameters, $Q1$ and $Q2$ 
(see Figs. \ref{q}, \ref{q2}, and \ref{figq1q2}), in which we identify
color-windows mostly populated by  RSG stars and massive  windy hot stars (e.g. WRs). 
Contamination by other stars can be reduced by taking into account luminosity and 
distance information.

The  $Q2$ values is based on near and mid-infrared data, which are not taken simultaneously.
In order to reduce the uncertainty due to non simultaneity,  we  use
information on variability, e.g., flags provided in the MSX catalog,
comparison of the 8 \um\ measurements from GLIMPSE and MSX \citep[e.g.][]{robitaille07}, 
and of the $J$ and \Ks\ measurements from 2MASS and DENIS \Ks\  \citep[e.g.][]{messineo04,schultheis01}.

In the following, we list our new proposed classification steps.
 
\begin{itemize}
\item  With available  $J, H, $\Ks\  magnitudes from 2MASS and
8.0\um\  magnitudes from GLIMPSE or MSX, we calculate the $Q1$ and $Q2$ parameters. 

\item We classify  as late-types those stars located above the relation
$Q1 = -0.20 \times (K_{S}-[8.0])+0.34$ mag (see Fig.\ \ref{q}), and we classify as 
early-type stars those stars below it. This relation applies only to stars with \Ks$-[8.0] <4$ mag.

\item 
To explore the impact of each constraint, we select candidate early-type stars 
with free-free emission in several manners. We call "free-1" stars those early-type 
stars with $Q2 < -1$ mag, and "free-2" stars with the additional condition 
$Q2 > 11.25 \times Q1 -2.38$ mag. We call "free-3"  stars with  extra  additional conditions,
the GLIMPSE criteria  specified by \citet{hadfield07,mauerhan11}.
We compare the new criteria with those used by \citet[hereafter, called "free-MVM11"][]{mauerhan11}.

\item 
We select  late-type stars with $0.1< Q1 < 0.5$ mag and $-1.1 < Q2 <  1.5$ as candidate RSGs (cRSG1),
or more stringently with  $0.1< Q1 < 0.5$ mag and $+0.5 < Q2 <  1.5$ mag (cRSG2).
When using the additional condition $[3.6]-[4.5] < -0.4$ mag, a detection rate of 70\%
is expected (see Fig. \ref{q2}). We call this more restricted 
color selection as cRSG3. We do not consider candidate RSGs with an indication of variability.

\item The degeneracy  between candidate RSGs and AGB-SRs  can be removed by additional information 
on luminosity, and therefore extinction and  distance. We suggest to calculate the total extinction 
(circumstellar and interstellar) as described in \citet{messineo05}, and to use the model of 
Galactic dust distribution by \citet{drimmel03} for estimating distances. With bolometric corrections, 
distances, and extinction estimates, luminosities can be derived. 

\end{itemize}

For each of the samples, we  ignored  the a priori knowledge on spectral types, 
and calculated the fractions of retrieved free-free emitters and RSGs, by using
the categories defined above.  The results of our experiments are  listed in 
Table \ref{result}. 
     
\begin{table*}
\caption{\label{result} Fractions of retrieved types.}
\begin{tabular}{lllllllllll}
Type     &Early-type& free-1& free-2& free-3 &free-MVM11 &Late-type &cRSG1&cRSG2& cRSG3\\
\hline
 WR all &   0.86 &   0.85 &   0.85 &   0.78 &   0.80 &   0.14 &   0.02 &   0.00 &   0.00\\
WR MVM1 &   0.75 &   0.73 &   0.73 &   0.72 &   0.73 &   0.25 &   0.00 &   0.00 &   0.00\\
WR SFZ1 &   0.90 &   0.90 &   0.90 &   0.90 &   0.90 &   0.10 &   0.05 &   0.00 &   0.00\\
   LBV  &   0.67 &   0.67 &   0.67 &   0.33 &   0.44 &   0.33 &   0.11 &   0.00 &   0.00\\
   RSG  &   0.11 &   0.00 &   0.00 &   0.00 &   0.00 &   0.89 &   0.72 &   0.42 &   0.22\\
 OH/IR  &   0.52 &   0.33 &   0.15 &   0.00 &   0.15 &   0.48 &   0.04 &   0.00 &   0.00\\
   SiO  &   0.25 &   0.16 &   0.16 &   0.01 &   0.19 &   0.75 &   0.04 &   0.00 &   0.00\\
AGB-LAV &   0.03 &   0.01 &   0.01 &   0.01 &   0.03 &   0.97 &   0.36 &   0.06 &   0.00\\
AGB-SR  &   0.37 &   0.02 &   0.02 &   0.00 &   0.00 &   0.63 &   0.58 &   0.35 &   0.00\\
\hline
\end{tabular}
\tablefoot{
 For each sample of star, we list the fraction of retrieved early-type (Early-type),
three estimates of the fractions of free-free emitters (free-1, free-2, free-3, and free-MVM11), the fraction
of retrieved late-type stars (Late-type), and three estimates of RSG stars (cRSG1, cRSG2, cRSG3, and cRSG4).
Only stars with counterparts in the three 2MASS and four GLIMPSE bands are considered.
For WR stars, we consider the whole sample  
alltogether (WR all),  as well as the newly detected samples by \citet[WR MVM11][]{mauerhan11} 
and \citet[WR SFZ11][]{shara11}. The selection of free-free emitters is applied only to 
early-type stars; the selection of RSG stars is applied only to late-type stars.}
\end{table*}

The selection of early and late-type stars retrieves 85\%  of WR stars from \citet{vanderhucht01}, 
but only 63\% of  SR stars, and 48\% of  the OH/IR stars listed by \citet{alard01}.

The Q2 parameter is a good discriminant for selecting free-free emitters ("free-1"); 
additional constraints on GLIMPSE colors are needed to reduce contamination 
by late-type stars ("free-3"). The pre-selection of early and late-type stars allows retrieval of 
 75\% of free-free emitters, but strongly reduces the contamination by AGBs; 
the number of AGB stars erroneously retrieved as free-free emitters decreases from 
33\% ("free-1") to a few percent ("free-3").   
For a  comparison with \citet{mauerhan11}, we also use their 
selection criteria \citep[Fig. 1 in ][]{mauerhan11}, which we  label as free-MVM11.
The Q1-Q2 based criteria provide  a  detection efficiency comparable to that by
Mauerhan et al. on the whole sample of WRs.
These authors applied additional criteria than the color-color criteria by \citet{hadfield07},
based on the \Ks\ versus $J-$\Ks\ diagram. We also reclassified separately their sample of 61 WRs, which
is composed of more distant and  obscured WR stars with our classification scheme, and retrieved 
75\% of them as early-type stars, and 73\% as free-free emitters (free-1). 

The selection of a narrow window in the $Q1$ versus $Q2$ plane with the additional limits on GLIMPSE 
colors allows selecting  bonafide RSG stars.

\section{Summary and discussion}
\subsection{Summary}
We analyzed 2MASS and GLIMPSE color properties of samples of Galactic evolved stars.
Samples of O-rich AGB stars (SRs, Miras, OH/IR stars, and SiO masing stars), RSG stars, 
WRs, and LBVs  were collected from existing literature. Several color-color diagrams were 
analyzed, aiming to identify the best combination for stellar photometric classifications. 

The $J-H$ versus $H-$\Ks\ diagram, and the $J-$\Ks\ versus  \Ks$-[8.0]$ diagram are particular 
useful. In these planes, the colors of WR stars  and mass-losing late type stars 
deviate from the interstellar reddening  vector.  WRs have a mid-infrared excess 
due to their free-free emission. Late-type stars often have mid-infrared excess due to
their dusty envelopes.  The deviation from the reddening vector in the \Ks$-[8.0]$ versus 
$J-$\Ks\ plane increases going from SRs to Miras and OH/IR stars, i.e. with increasing mass-loss rates. 

The $J-H$ versus $H-$\Ks, and the $J-$\Ks\ versus  \Ks$-[8.0]$ diagrams together can be used 
to discriminate between interstellar and circumstellar reddening, and identify different types 
of stars with envelopes. Stellar classification is hampered by color overlaps
between sources of different types, and scatter due to variability and non-simultaneity of 
the observations. Special color windows have been identified for locating  WR and RSG stars.

We investigated two extinction free parameters, $Q1$ and $Q2$. $Q1$ is a measure of  the deviation 
from the reddening vector in the  $J-H$ versus $H-$\Ks\  plane \citep{negueruela07}.
$Q2$ is a newly defined parameter, which measures  the deviation from the reddening vector 
in the $J-$\Ks\ versus  \Ks$-[8.0]$ plane. 
The $Q1$ and $Q2$ parameters allow for an efficient selection of stars with free-free
emission (e.g. WR stars), and candidate RSG stars.
Stars  with $Q1 < -1.0$ mag and $Q2 > 11.25 \times Q1 -2.38$ mag are candidate hot 
massive stars with free-free radiation.
A large number ($\sim 40\%$) of Galactic RSGs are found  with $0.1 < Q1 < 0.5$ mag, 
and  $-1.1 < Q2 < 1.5$ mag. 
Selections can be further improved with additional conditions (e.g. with the $[3.6]-[4.5]$ color).

A combination of mid- and near-infrared measurements allows to statistically distinguish  
between evolved mass-losing stars and early-type stars. 

\subsection{Selection of free-free emitters}

We have found a clear separation between massive stars with free-free emission (e.g. WR stars)  
and late-type stars in the $Q1$ versus \Ks$-[8.0]$ diagram, and in the $Q1$ versus $Q2$ diagram. 

A combination of   $[3.6]-[8.0]$ versus $[3.6]-[4.5]$  and \Ks$-[8.0]$  versus $J-$\Ks\ diagrams
(Figs. \ref{lucy1} and \ref{lucy2}) were chosen by \citet{hadfield07} to best identify WR stars
from other field stars. 
Recently, \citet{mauerhan11} revised the selection of WR stars 
with additional constrains on the  $H-$\Ks\ versus $J-$\Ks\ diagram, and \Ks\ versus $J-$\Ks\ diagram, 
to obtained a spectroscopic detection rate of 95\%  in early-type stars, and 20\% in WR stars.
The criteria suggested by \citet{hadfield07} and \citet{mauerhan11} depend on interstellar 
extinction, and suffer some contamination by dust enshrouded AGBs. 
Our new selection based on $Q1$ and $Q2$ has the advantage of simplicity, based on a single 
diagram that is independent of interstellar  extinction. 
A fraction of WRs (about  15 \% ), however,  are found in the region of late-type stars, 
and  are missed by our selection of early type stars -- among them several WC stars, which are  
known to have dusty envelopes. 

The complex structures of LBV envelopes, which  are dusty and often extended, result in a broad 
range of colors. $Q2$ varies from $-2.2$ to $-0.7$ mag and $Q1$ from $-$0.34 to 0.0 mag. LBVs 
occupy a broad color spaces, and  cannot  be identified with photometry of point sources alone.

\subsection{Pulsation and stellar colors}

A correlation is newly found  between pulsation types and GLIMPSE colors.
\citet{alvarez00} writes:
"Pulsation produces much more extended atmospheres, and in addition dense cool
layers may result from the periodic outwards running shocks.
In various ways pulsation thus leads to the existence of regions
where relatively low temperatures  are combined with relatively high densities, 
conditions that favor the formation of water, and dust."

For AGBs, we have located   a sequence of increasing $[3.6]-[8.0]$ colors with increasing 
pulsation amplitudes (from SRs to OH/IR stars). The behaviors of the $[3.6]-[4.5]$ color 
with spectral types is more complex.  Generally, SRs and LAVs (without SiO maser) 
show a nice correlation of increasing $[3.6]-[4.5]$ colors with increasing  $[3.6]-[8.0]$ 
colors, but the subgroup of  SiO masing stars appears  to be  more scattered and to have bluer 
$[3.6]-[4.5]$ colors. In the Appendix, a few ISO-SWS representative spectra of evolved 
late-type stars are shown. The increase in $[3.6]-[4.5]$ colors going from SRs to 
LAVs is due  to strong continuum absorption by water in the GLIMPSE 3.6 \um\ band \citep{matsuura02}.
In contrast, the bluer color of masing AGB stars is likely due to a combination of stronger
water absorption  (3.6 \um\ band), and presence of absorption due to SiO and CO$_2$ molecules
in the 4.5 \um\ band. 

Several  parameters, such as luminosity,  rotation, turbulence, and metallicity, 
play an important role in determining the dust chemistry of RSG envelopes.
RSG stars with regular pulsation  display redder $[3.6]-[4.5]$ color than irregular RSGs,
as suggested by existing ISO-SWS spectra (see Appendix). Pulsation appears, therefore,
a key parameter for mass-loss rates also in RSG stars.
The GLIMPSE  $[3.6]-[4.5]$ and $[3.6]-[8.0]$ colors of RSGs correlate with mass-loss 
rates (see Appendix).  RSGs show a broad range of $[3.6]-[4.5]$ colors, and can be bluer than AGB stars.
Photometric monitoring of Galactic RSGs, as well as follow-up mid-infrared spectroscopic  
observations of the 3 \um\ region, is needed to further explore a correlation 
between  chemistry and  pulsation properties in RSGs,  as suggested by  Fig.\ \ref{lucy1}. 
The RSGCs share the same location in the Galaxy, and  likely have similar metallicity
\citep{davies09}, but their RSG members have a large spread in $[3.6]-[4.5]$ colors.
This suggests that pulsation affect the mid-infrared properties of RSGs
more than metallicity.

\subsection{Selection of RSG stars}
We confirm that the selection of RSGs based on $Q1$ values adopted by \citet{clark09},
\citet{negueruela10}, and \citet{negueruela11} (as remarked already in these works) excludes 
possible dust enshrouded RSG stars, and its strongly contaminated by AGBs (SRs and Miras).
Selection of  bonafide RSG stars solely based on the $Q1$ parameter 
is impossible. The $Q1$ parameter remains, however, a valid tool to select
clusters with RSGs. RSGs  are  often found in massive clusters, because of  their young 
ages ($<30$ Myr) \citep[e.g.][]{figer06,davies07}. Masing AGBs are typically found 
in isolation \citep[e.g.][]{messineo02, sevenster02,sjouwerman98}. 

We propose a selection of candidate RSG stars based on a larger set of constraints:
the $Q1$ and $Q2$ values, variability information, $[3.6]-[4.5]$ colors,
and apparent magnitudes.
We can distinguish between interstellar and circumstellar extinction, using the $Q1$ and $Q2$
parameters. First distance estimates
can be obtained by assuming an interstellar-extinction versus distance relation \citep{drimmel03};
stellar luminosities can, then, be estimated by using adeguate bolometric corrections.

\subsection{Final remarks}

While our new criteria do not allow classification of all IR bright  Galactic stars detected by 
GLIMPSE or MSX, the number of known WR stars and RSGs can highly be increased. 
The proposed criteria are particularly 
useful for studying stellar populations, for which an average extinction and distance can be assumed. 
They will allow to efficiently detect massive stars  in  giant HII regions and supernova remnants.

The presented  datasets open new possibilities for addressing, 
and pose questions concerning our understanding of stellar reddening,
bolometric corrections, and extinction ratios.

\begin{appendix}
\begin{figure*}[!]
\begin{centering}
\resizebox{0.49\hsize}{!}{\includegraphics[angle=0]{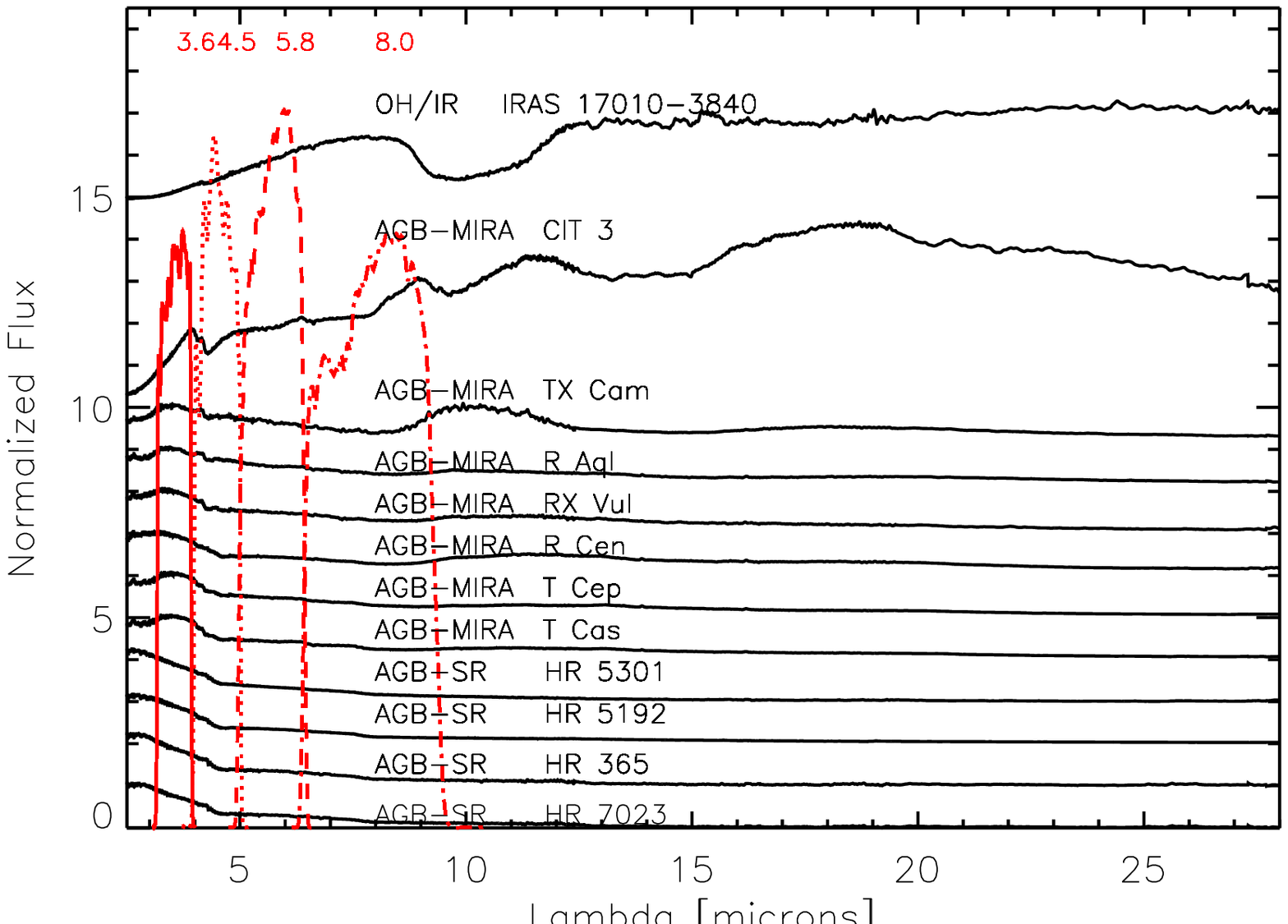}}
\resizebox{0.49\hsize}{!}{\includegraphics[angle=0]{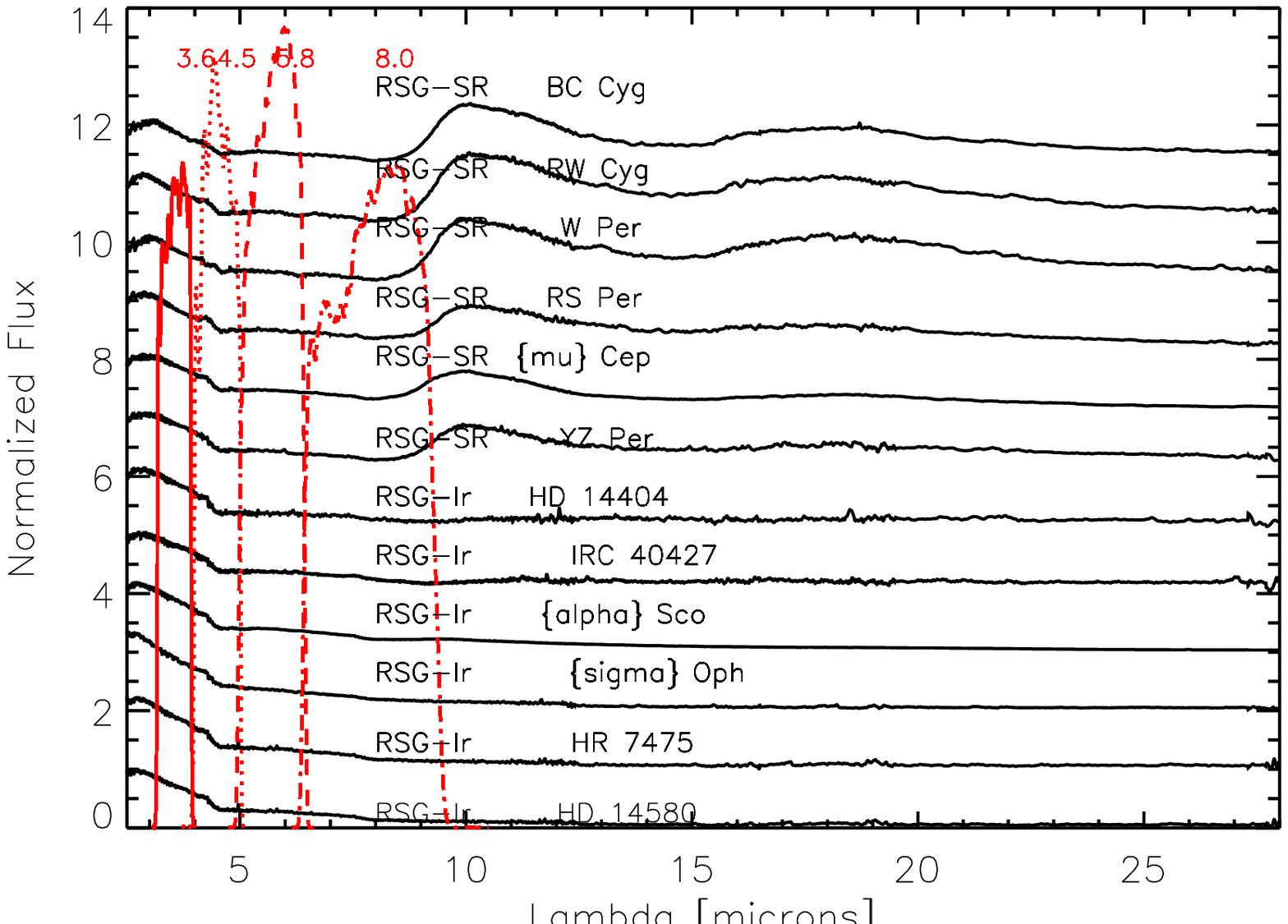}}
\end{centering}
\caption{\label{plotspectra}  Left: A representative samples of ISO-SWS spectra of AGB stars.
Over-plotted transmission curves of the four SPITZER/IRAC filters.
 Right: A representative samples of ISO-SWS spectra of RSG stars.
Over-plotted transmission curves of the four SPITZER/IRAC filters.}
\end{figure*}

\begin{figure*}[!]
\begin{centering}
\resizebox{0.7\hsize}{!}{\includegraphics[angle=0]{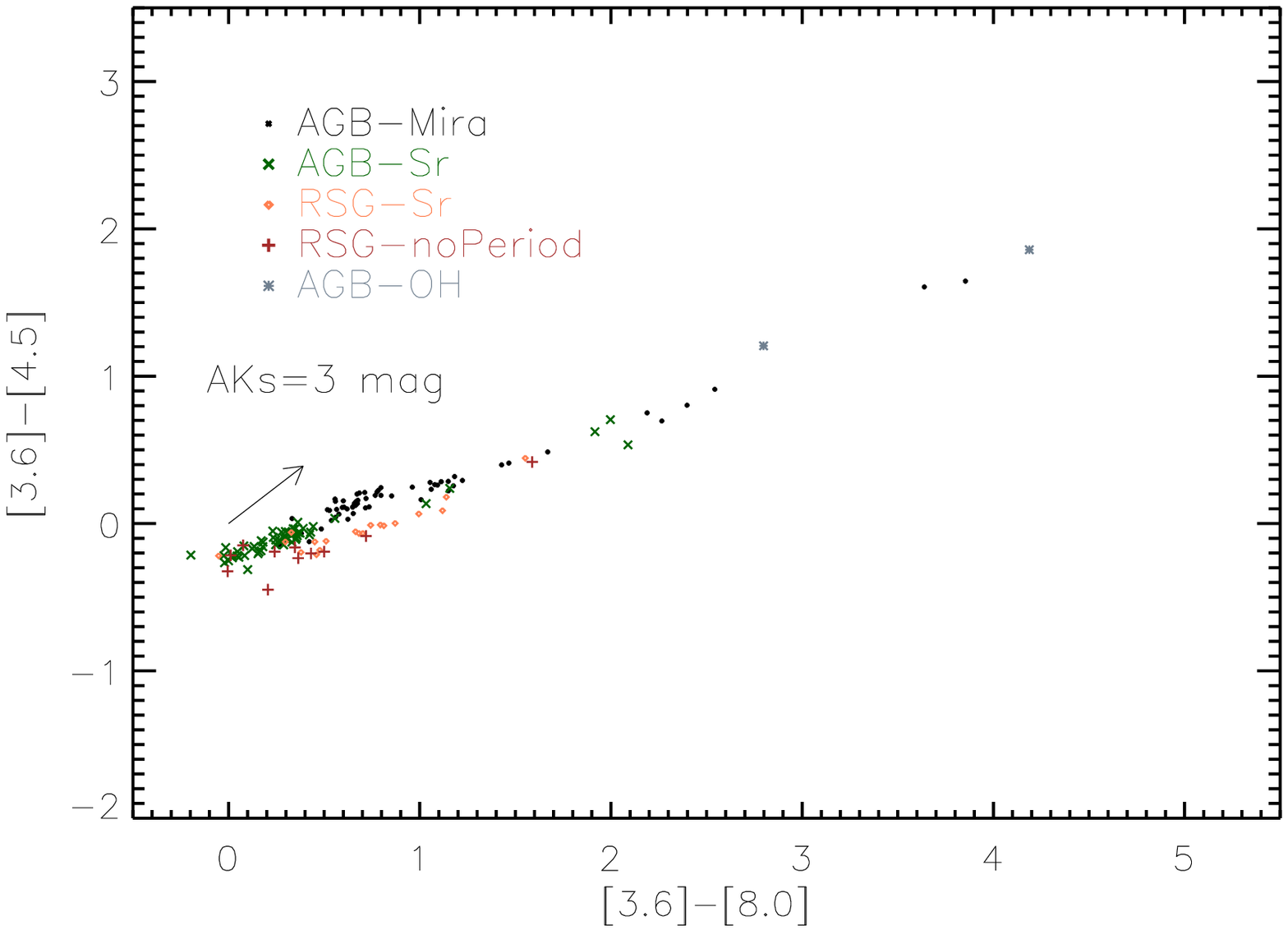}}
\end{centering}
\caption{\label{synt} Synthetic GLIMPSE $[3.6]-[8.0]$ versus $[3.6]-[4.5]$ diagram
of evolved late-type stars observed with ISO-SWS \citep{sloan03}.
AGB-Miras are shown with black dots, AGB-SR with crosses, RSG-SR with diamonds,
RSG irregular with plus signs, and OH/IR stars with starred symbols.
The arrow indicates the direction of the reddening vector following the extinction 
ratios derived by \citet{indebetouw05}.}
\end{figure*}

\begin{figure*}[!]
\begin{centering}
\resizebox{0.49\hsize}{!}{\includegraphics[angle=0]{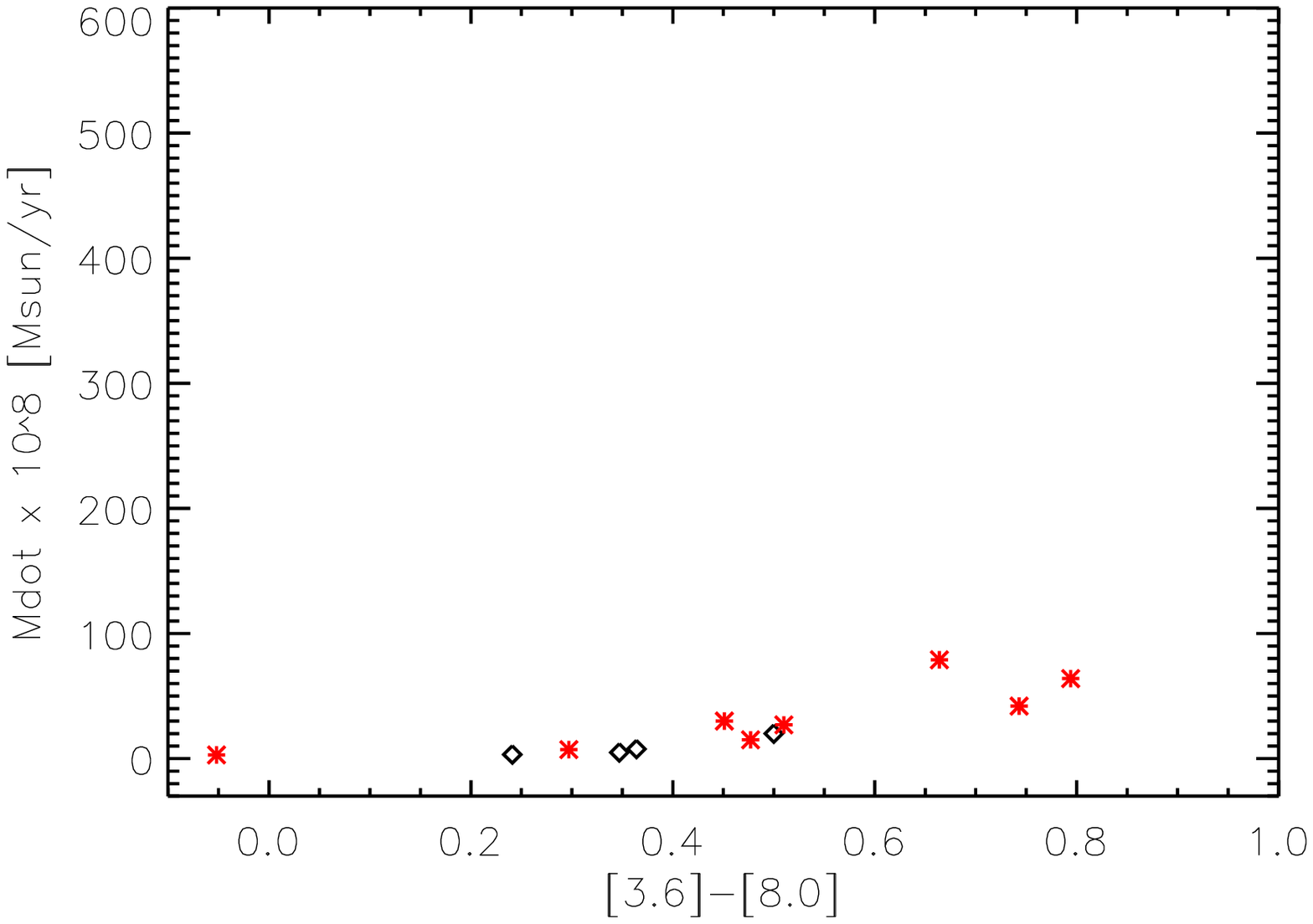}}
\resizebox{0.49\hsize}{!}{\includegraphics[angle=0]{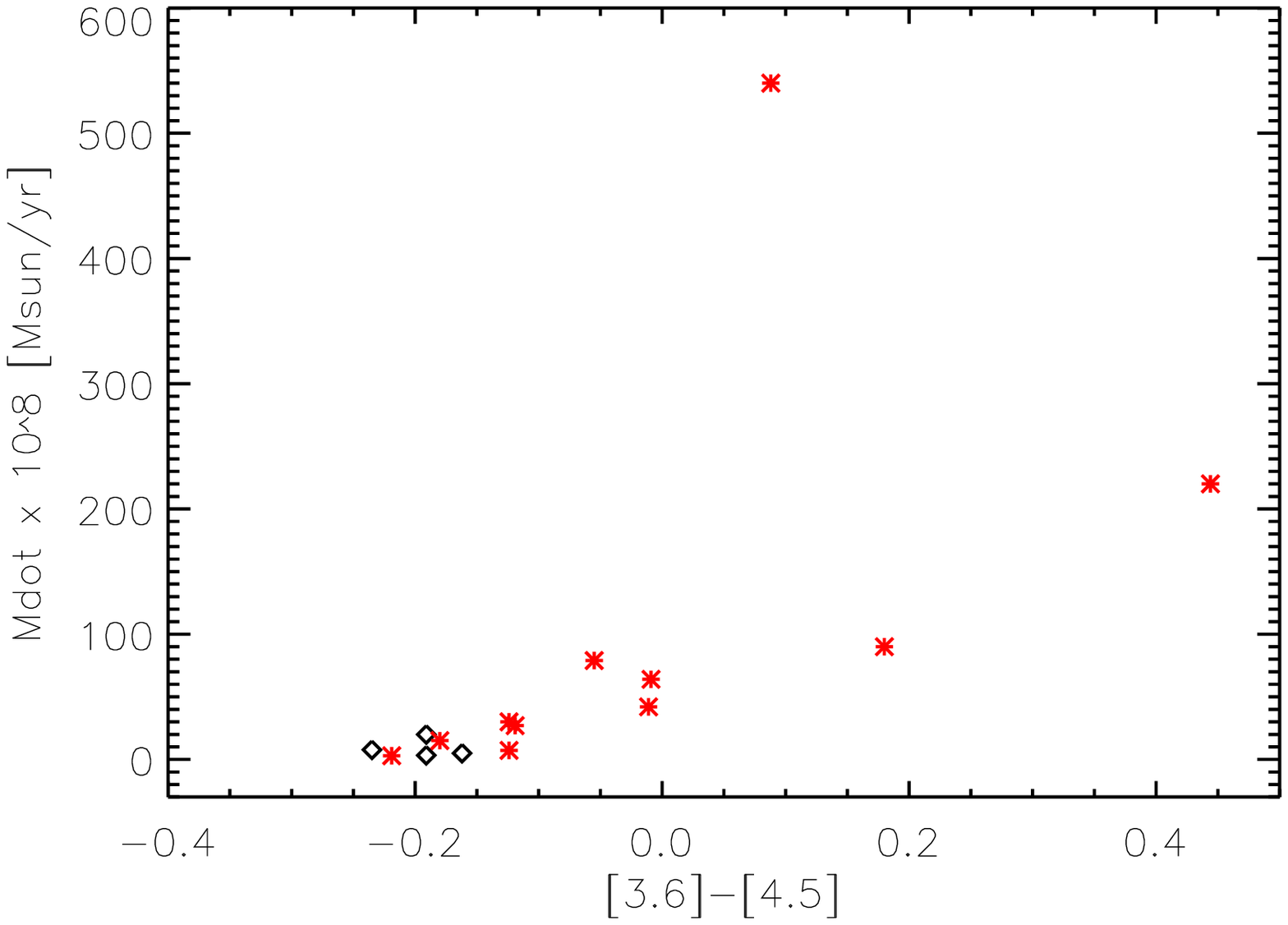}}
\end{centering}
\caption{\label{massloss}  Right: Synthetic $[3.6]-[4.5]$ of RSG stars versus mass-loss rates.
Magnitudes are obtained  from ISO-SWS spectra \citep{sloan03}, and mass-loss rates are given by
\citet{verhoelst09}. Starred symbols are RSG stars with periodic pulsations (SR), while diamonds are
irregular RSG stars.  Left: Synthetic $[3.6]-[8.0]$ of RSG stars modeled by 
\citet{verhoelst09} versus their inferred mass-loss rates. Starred symbols are 
RSG stars with periodic pulsations (SR), while diamonds are
irregular RSG stars.}
\end{figure*}

\section{ISO-SWS spectra of late-type stars}

Molecular absorption and/or emission and dust features    affect
the $[3.6]-[4.5]$ and $[3.6]-[8.0]$  colors.
In order to understand the observed trend of increasing $[3.6]-[4.5]$ 
and $[3.6]-[8.0]$ colors of late-type stars with pulsation types, we analysed  
ISO-SWS spectra  (see Fig. \ref{plotspectra}).
From the  flux-calibrated  ISO-SWS library  of \citet{sloan03}, we selected  spectra of stars 
with known spectral types and variabilities: AGB-Miras, AGB-SRs, and OH/IR stars, and RSGs.

Water, SiO, CO$_2$, and CO absorption  may be present in  the  3.0 to  5.0 \um\ spectral region
of late-type stars \citep{sylvester99,matsuura02,cami02}. ISO-SWS spectra of RSGs and AGB-SRs are 
remarkably different from those of AGB Miras 
(see Fig.\ \ref{plotspectra} ). 
Typically RSGs and AGB-SRs do not have strong molecular signatures in between 3 and 4 \um,
and display a decreasing flux in this region (GLIMPSE 3.6 \um). 
AGB-Miras are dominated by water, SiO, CO$_2$, and CO absorption, 
and the resulting  spectral flux  between 3 and 4 \um\ can have a positive slope. 
 This is attributed to a different "MOLspheres",  layers of molecular material in an 
extended atmosphere (Miras), where water is the main source of opacity 
\citep{tsuji97,matsuura02,verhoelst09}. 
The strength of water absorption  varies  with stellar pulsation phase \citep{matsuura02}.  

\citet{verhoelst09}  explains the continuum shape of RSGs in the 2.5-3 \um\ regions with an 
extra dust component (metallic Fe, amorphous C, or O-rich micron-sized grains) or with
chromospheric emission. The similarity of the continuum  of AGB SRs and RSG stars 
in the 2.5 and 3 \um\ region   suggests that the attribution of extra continuum emission  
to free-free radiation from chromospheric activities is less likely.
Regularly pulsating RSGs have larger mass-loss (as suggested
by the strength of the silicate emission at 9.7 \um) than  RSGs with irregular pulsations
(Fig.\ \ref{plotspectra}).

The ISO-SWS flux densities at the effective wavelength of the GLIMPSE filters 
were measured, and flux densities were converted into magnitudes  using the GLIMPSE zero-points. 
The synthetic  $[3.6]-[4.5]$ versus $[3.6]-[8.0]$ diagram is shown in Fig.\ \ref{synt}, and 
well reproduces the sequence of increasing  $[3.6]-[4.5]$ with $[3.6]-[8.0]$ colors.
A clear sequence of redder colors is seen in AGB stars, going from SRs, to Miras and OH/IR stars.
An increase in the $[3.6]-[8.0]$ color is mostly due to an increase of  silicate
emission at 9.7 \um, which for optically thin envelopes well correlates with mass-loss.
Their  $[3.6]-[4.5]$ colors are redder than those of  AGB SR stars (Figs.\ \ref{plotspectra}),
because in the spectra of Miras there is strong water absorption.  
RSGs  and AGB SR stars have  weaker water absorption than AGB Miras.
\citet{yang11} suggest that RSGs have bluer $[3.6]-[4.5]$  due to a continuum depression 
around 4.5 \um\ by CO bands. RSG-SR stars have redder  $[3.6]-[4.5]$ colors 
than RSGs with irregular pulsations. Only weak water absorption is seen in some 
variable RSG stars.

 In Fig. \ref{massloss}, we plot our synthetic GLIMPSE colors of RSG stars 
versus  mass-loss rates, as inferred by \citet{verhoelst09}.
Both $[3.6]-[4.5]$ and  $[3.6]-[8.0]$ colors correlate with mass-loss rates.
The correlation of mass loss with the $[3.6]-[8.0]$ is explained by the increasing 
silicate strength at 9.7 \um.  The correlation of mass loss with  $[3.6]-[4.5]$ color
can be due to CO at $4.3$ \um, and/or to the existence of an extra source of opacity (dust component), 
as suggested by \citet{verhoelst09}.

\end{appendix}

\begin{acknowledgements}

This publication makes use of data products from the Two Micron All Sky Survey, which is a joint project 
of the University of Massachusetts and the Infrared Processing and Analysis Center/California Institute of Technology, 
funded by the National Aeronautics and Space Administration and the National Science Foundation.
This work is based [in part] on observations made with the Spitzer Space Telescope, 
which is operated by the Jet Propulsion Laboratory, California Institute of Technology under a contract with NASA.
The DENIS project was supported, in France by the Institut National des Sciences de l'Univers, the Education Ministry 
and the Centre National de la Recherche Scientifique, in Germany by the State of Baden-W�rtemberg, in Spain by the 
DGICYT, in Italy by the Consiglio Nazionale delle Ricerche, in Austria by the Fonds zur F�rderung der 
wissenschaftlichen Forschung and the Bundesministerium fuer Wissenschaft und Forschung. 
This research made use of data products from the
Midcourse Space Experiment, the processing of which was funded by the Ballistic Missile Defence Organization with additional
support from the NASA office of Space Science. This research has made use of the SIMBAD data base, operated at CDS, Strasbourg,
France. This work was funded by the ERC Advanced Investigator Grant GLOSTAR (247078).
MM thanks  Jon Mauerhan for sending her the "broad" color criterion for WR stars.
MM  is grateful to Frank Bertoldi, Don Figer, Vasilii Gvaramadze, 
Peter-Rolf Kudritzki, and  Maria Massi for stimulating discussions on massive stars.
We thank the referee Guandanlini Roald for his careful reading of our manuscript.

\end{acknowledgements}


\begin{thebibliography}{118}
\expandafter\ifx\csname natexlab\endcsname\relax\def\natexlab#1{#1}\fi

\bibitem[{{Abraham} {et~al.}(2005){Abraham}, {Falceta-Gon{\c c}alves},
  {Dominici}, {Caproni}, \& {Jatenco-Pereira}}]{abraham05}
{Abraham}, Z., {Falceta-Gon{\c c}alves}, D., {Dominici}, T., {Caproni}, A., \&
  {Jatenco-Pereira}, V. 2005, \mnras, 364, 922

\bibitem[{{Alard} {et~al.}(2001){Alard}, {Blommaert}, {Cesarsky}, {Epchtein},
  {Felli}, {Fouque}, {Ganesh}, {Genzel}, {Gilmore}, {Glass}, {Habing}, {Omont},
  {Perault}, {Price}, {Robin}, {Schultheis}, {Simon}, {van Loon}, {Alcock},
  {Allsman}, {Alves}, {Axelrod}, {Becker}, {Bennett}, {Cook}, {Drake},
  {Freeman}, {Geha}, {Griest}, {Lehner}, {Marshall}, {Minniti}, {Nelson},
  {Peterson}, {Popowski}, {Pratt}, {Quinn}, {Sutherland}, {Tomaney},
  {Vandehei}, \& {Welch}}]{alard01}
{Alard}, C., {Blommaert}, J.~A.~D.~L., {Cesarsky}, C., {et~al.} 2001, \apj,
  552, 289

\bibitem[{{Alcock} {et~al.}(1999){Alcock}, {Allsman}, {Alves}, {Axelrod},
  {Becker}, {Bennett}, {Cook}, {Drake}, {Freeman}, {Geha}, {Griest}, {Lehner},
  {Marshall}, {Minniti}, {Peterson}, {Popowski}, {Pratt}, {Nelson}, {Quinn},
  {Stubbs}, {Sutherland}, {Tomaney}, {Vandehei}, {Welch}, \& {The MACHO
  Collaboration}}]{alcock99}
{Alcock}, C., {Allsman}, R.~A., {Alves}, D.~R., {et~al.} 1999, \pasp, 111, 1539

\bibitem[{{Alvarez} {et~al.}(2000){Alvarez}, {Lan{\c c}on}, {Plez}, \&
  {Wood}}]{alvarez00}
{Alvarez}, R., {Lan{\c c}on}, A., {Plez}, B., \& {Wood}, P.~R. 2000, \aap, 353,
  322

\bibitem[{{Barniske} {et~al.}(2008){Barniske}, {Oskinova}, \&
  {Hamann}}]{barniske08}
{Barniske}, A., {Oskinova}, L.~M., \& {Hamann}, W. 2008, \aap, 486, 971

\bibitem[{{Benjamin} {et~al.}(2003){Benjamin}, {Churchwell}, {Babler}, {Bania},
  {Clemens}, {Cohen}, {Dickey}, {Indebetouw}, {Jackson}, {Kobulnicky},
  {Lazarian}, {Marston}, {Mathis}, {Meade}, {Seager}, {Stolovy}, {Watson},
  {Whitney}, {Wolff}, \& {Wolfire}}]{benjamin03}
{Benjamin}, R.~A., {Churchwell}, E., {Babler}, B.~L., {et~al.} 2003, \pasp,
  115, 953

\bibitem[{{Bernabei} \& {Polcaro}(2001)}]{bernabei01}
{Bernabei}, S. \& {Polcaro}, V.~F. 2001, \aap, 366, 817

\bibitem[{{Bibby} {et~al.}(2008){Bibby}, {Crowther}, {Furness}, \&
  {Clark}}]{bibby08}
{Bibby}, J.~L., {Crowther}, P.~A., {Furness}, J.~P., \& {Clark}, J.~S. 2008,
  \mnras, 386, L23

\bibitem[{{Blum} {et~al.}(2003){Blum}, {Ram{\'{\i}}rez}, {Sellgren}, \&
  {Olsen}}]{blum03}
{Blum}, R.~D., {Ram{\'{\i}}rez}, S.~V., {Sellgren}, K., \& {Olsen}, K. 2003,
  \apj, 597, 323

\bibitem[{{Bonanos} {et~al.}(2010){Bonanos}, {Lennon}, {K{\"o}hlinger}, {van
  Loon}, {Massa}, {Sewilo}, {Evans}, {Panagia}, {Babler}, {Block}, {Bracker},
  {Engelbracht}, {Gordon}, {Hora}, {Indebetouw}, {Meade}, {Meixner}, {Misselt},
  {Robitaille}, {Shiao}, \& {Whitney}}]{bonanos10}
{Bonanos}, A.~Z., {Lennon}, D.~J., {K{\"o}hlinger}, F., {et~al.} 2010, \aj,
  140, 416

\bibitem[{{Bressan} {et~al.}(1984){Bressan}, {Chiosi}, \&
  {Bertelli}}]{bressan84}
{Bressan}, A.~G., {Chiosi}, C., \& {Bertelli}, G. 1984, \aap, 130, 279

\bibitem[{{Buchanan} {et~al.}(2006){Buchanan}, {Kastner}, {Forrest}, {Hrivnak},
  {Sahai}, {Egan}, {Frank}, \& {Barnbaum}}]{buchanan06}
{Buchanan}, C.~L., {Kastner}, J.~H., {Forrest}, W.~J., {et~al.} 2006, \aj, 132,
  1890

\bibitem[{{Buchanan} {et~al.}(2009){Buchanan}, {Kastner}, {Hrivnak}, \&
  {Sahai}}]{buchanan09}
{Buchanan}, C.~L., {Kastner}, J.~H., {Hrivnak}, B.~J., \& {Sahai}, R. 2009,
  \aj, 138, 1597

\bibitem[{{Cami}(2002)}]{cami02}
{Cami}, J. 2002, PhD thesis, University of Amsterdam

\bibitem[{{Caron} {et~al.}(2003){Caron}, {Moffat}, {St-Louis}, {Wade}, \&
  {Lester}}]{caron03}
{Caron}, G., {Moffat}, A.~F.~J., {St-Louis}, N., {Wade}, G.~A., \& {Lester},
  J.~B. 2003, \aj, 126, 1415

\bibitem[{{Churchwell} {et~al.}(2009){Churchwell}, {Babler}, {Meade},
  {Whitney}, {Benjamin}, {Indebetouw}, {Cyganowski}, {Robitaille}, {Povich},
  {Watson}, \& {Bracker}}]{churchwell09}
{Churchwell}, E., {Babler}, B.~L., {Meade}, M.~R., {et~al.} 2009, \pasp, 121,
  213

\bibitem[{{Clark} {et~al.}(2003){Clark}, {Egan}, {Crowther}, {Mizuno},
  {Larionov}, \& {Arkharov}}]{clark03}
{Clark}, J.~S., {Egan}, M.~P., {Crowther}, P.~A., {et~al.} 2003, \aap, 412, 185

\bibitem[{{Clark} {et~al.}(2005){Clark}, {Larionov}, \& {Arkharov}}]{clark05}
{Clark}, J.~S., {Larionov}, V.~M., \& {Arkharov}, A. 2005, \aap, 435, 239

\bibitem[{{Clark} {et~al.}(2009){Clark}, {Negueruela}, {Davies}, {Larionov},
  {Ritchie}, {Figer}, {Messineo}, {Crowther}, \& {Arkharov}}]{clark09}
{Clark}, J.~S., {Negueruela}, I., {Davies}, B., {et~al.} 2009, \aap, 498, 109

\bibitem[{{Cohen} {et~al.}(1975){Cohen}, {Kuhi}, \& {Barlow}}]{cohen75}
{Cohen}, M., {Kuhi}, L.~V., \& {Barlow}, M.~J. 1975, \aap, 40, 291

\bibitem[{{Comer{\'o}n} {et~al.}(2004){Comer{\'o}n}, {Torra}, {Chiappini},
  {Figueras}, {Ivanov}, \& {Ribas}}]{comeron04}
{Comer{\'o}n}, F., {Torra}, J., {Chiappini}, C., {et~al.} 2004, \aap, 425, 489

\bibitem[{{Conti} {et~al.}(1995){Conti}, {Hanson}, {Morris}, {Willis}, \&
  {Fossey}}]{conti95}
{Conti}, P.~S., {Hanson}, M.~M., {Morris}, P.~W., {Willis}, A.~J., \& {Fossey},
  S.~J. 1995, \apjl, 445, L35

\bibitem[{{Cutri} {et~al.}(2003){Cutri}, {Skrutskie}, {van Dyk}, {Beichman},
  {Carpenter}, {Chester}, {Cambresy}, {Evans}, {Fowler}, {Gizis}, {Howard},
  {Huchra}, {Jarrett}, {Kopan}, {Kirkpatrick}, {Light}, {Marsh}, {McCallon},
  {Schneider}, {Stiening}, {Sykes}, {Weinberg}, {Wheaton}, {Wheelock}, \&
  {Zacarias}}]{cutri03}
{Cutri}, R.~M., {Skrutskie}, M.~F., {van Dyk}, S., {et~al.} 2003, {2MASS All
  Sky Catalog of point sources.}

\bibitem[{{Davies} {et~al.}(2007){Davies}, {Figer}, {Kudritzki}, {MacKenty},
  {Najarro}, \& {Herrero}}]{davies07}
{Davies}, B., {Figer}, D.~F., {Kudritzki}, R., {et~al.} 2007, \apj, 671, 781

\bibitem[{{Davies} {et~al.}(2009{\natexlab{a}}){Davies}, {Figer}, {Kudritzki},
  {Trombley}, {Kouveliotou}, \& {Wachter}}]{davies09b}
{Davies}, B., {Figer}, D.~F., {Kudritzki}, R., {et~al.} 2009{\natexlab{a}},
  \apj, 707, 844

\bibitem[{{Davies} {et~al.}(2008){Davies}, {Figer}, {Law}, {Kudritzki},
  {Najarro}, {Herrero}, \& {MacKenty}}]{davies08}
{Davies}, B., {Figer}, D.~F., {Law}, C.~J., {et~al.} 2008, \apj, 676, 1016

\bibitem[{{Davies} {et~al.}(2009{\natexlab{b}}){Davies}, {Origlia},
  {Kudritzki}, {Figer}, {Rich}, {Najarro}, {Negueruela}, \& {Clark}}]{davies09}
{Davies}, B., {Origlia}, L., {Kudritzki}, R., {et~al.} 2009{\natexlab{b}},
  \apj, 696, 2014

\bibitem[{{Deguchi} {et~al.}(2004){Deguchi}, {Fujii}, {Glass}, {Imai}, {Ita},
  {Izumiura}, {Kameya}, {Miyazaki}, {Nakada}, \& {Nakashima}}]{deguchi04}
{Deguchi}, S., {Fujii}, T., {Glass}, I.~S., {et~al.} 2004, \pasj, 56, 765

\bibitem[{{Drimmel} {et~al.}(2003){Drimmel}, {Cabrera-Lavers}, \&
  {L{\'o}pez-Corredoira}}]{drimmel03}
{Drimmel}, R., {Cabrera-Lavers}, A., \& {L{\'o}pez-Corredoira}, M. 2003, \aap,
  409, 205

\bibitem[{{Egan} {et~al.}(2003){Egan}, {Price}, {Kraemer}, {Mizuno}, {Carey},
  {Wright}, {Engelke}, {Cohen}, \& {Gugliotti}}]{egan03}
{Egan}, M.~P., {Price}, S.~D., {Kraemer}, K.~E., {et~al.} 2003, VizieR Online
  Data Catalog, 5114, 0

\bibitem[{{Eggenberger} {et~al.}(2002){Eggenberger}, {Meynet}, \&
  {Maeder}}]{eggenberger02}
{Eggenberger}, P., {Meynet}, G., \& {Maeder}, A. 2002, \aap, 386, 576

\bibitem[{{Epchtein} {et~al.}(1999){Epchtein}, {Deul}, {Derriere},
  {Borsenberger}, {Egret}, {Simon}, {Alard}, {Balazs}, {de Batz}, {Cioni},
  {Copet}, {Dennefeld}, {Forveille}, {Fouque}, {Garzon}, {Habing}, {Holl},
  {Hron}, {Kimeswenger}, {Lacombe}, {Le Bertre}, {Loup}, {Mamon}, {Omont},
  {Paturel}, {Persi}, {Robin}, {Rouan}, {Tiphene}, {Vauglin}, \&
  {Wagner}}]{denis2}
{Epchtein}, N., {Deul}, E., {Derriere}, S., {et~al.} 1999, VizieR Online Data
  Catalog, 1, 2002

\bibitem[{{Fazio} {et~al.}(2004){Fazio}, {Hora}, {Allen}, {Ashby}, {Barmby},
  {Deutsch}, {Huang}, {Kleiner}, {Marengo}, {Megeath}, {Melnick}, {Pahre},
  {Patten}, {Polizotti}, {Smith}, {Taylor}, {Wang}, {Willner}, {Hoffmann},
  {Pipher}, {Forrest}, {McMurty}, {McCreight}, {McKelvey}, {McMurray}, {Koch},
  {Moseley}, {Arendt}, {Mentzell}, {Marx}, {Losch}, {Mayman}, {Eichhorn},
  {Krebs}, {Jhabvala}, {Gezari}, {Fixsen}, {Flores}, {Shakoorzadeh}, {Jungo},
  {Hakun}, {Workman}, {Karpati}, {Kichak}, {Whitley}, {Mann}, {Tollestrup},
  {Eisenhardt}, {Stern}, {Gorjian}, {Bhattacharya}, {Carey}, {Nelson},
  {Glaccum}, {Lacy}, {Lowrance}, {Laine}, {Reach}, {Stauffer}, {Surace},
  {Wilson}, {Wright}, {Hoffman}, {Domingo}, \& {Cohen}}]{fazio04}
{Fazio}, G.~G., {Hora}, J.~L., {Allen}, L.~E., {et~al.} 2004, \apjs, 154, 10

\bibitem[{{Felli} \& {Panagia}(1981)}]{felli81}
{Felli}, M. \& {Panagia}, N. 1981, \aap, 102, 424

\bibitem[{{Figer} {et~al.}(2006){Figer}, {MacKenty}, {Robberto}, {Smith},
  {Najarro}, {Kudritzki}, \& {Herrero}}]{figer06}
{Figer}, D.~F., {MacKenty}, J.~W., {Robberto}, M., {et~al.} 2006, \apj, 643,
  1166

\bibitem[{{Figer} {et~al.}(1999){Figer}, {McLean}, \& {Morris}}]{figer99}
{Figer}, D.~F., {McLean}, I.~S., \& {Morris}, M. 1999, \apj, 514, 202

\bibitem[{{Figer} {et~al.}(1997){Figer}, {McLean}, \& {Najarro}}]{figer97}
{Figer}, D.~F., {McLean}, I.~S., \& {Najarro}, F. 1997, \apj, 486, 420

\bibitem[{{Fritz} {et~al.}(2011){Fritz}, {Gillessen}, {Dodds-Eden}, {Lutz},
  {Genzel}, {Raab}, {Ott}, {Pfuhl}, {Eisenhauer}, \& {Yusef-Zadeh}}]{fritz11}
{Fritz}, T.~K., {Gillessen}, S., {Dodds-Eden}, K., {et~al.} 2011, \apj, 737, 73

\bibitem[{{Frogel} \& {Whitford}(1987)}]{frogel87}
{Frogel}, J.~A. \& {Whitford}, A.~E. 1987, \apj, 320, 199

\bibitem[{{Gehrz}(1989)}]{gehrz89}
{Gehrz}, R. 1989, in IAU Symposium, Vol. 135, Interstellar Dust, ed.
  {L.~J.~Allamandola \& A.~G.~G.~M.~Tielens}, 445

\bibitem[{{Georgelin} \& {Georgelin}(1976)}]{georgelin76}
{Georgelin}, Y.~M. \& {Georgelin}, Y.~P. 1976, \aap, 49, 57

\bibitem[{{Glass} {et~al.}(2001){Glass}, {Matsumoto}, {Carter}, \&
  {Sekiguchi}}]{glass01}
{Glass}, I.~S., {Matsumoto}, S., {Carter}, B.~S., \& {Sekiguchi}, K. 2001,
  \mnras, 321, 77

\bibitem[{{Groenewegen} \& {de Jong}(1993)}]{groenewegen93}
{Groenewegen}, M.~A.~T. \& {de Jong}, T. 1993, \aap, 267, 410

\bibitem[{{Gvaramadze} {et~al.}(2010){Gvaramadze}, {Kniazev}, {Fabrika},
  {Sholukhova}, {Berdnikov}, {Cherepashchuk}, \& {Zharova}}]{gvaramadze10}
{Gvaramadze}, V.~V., {Kniazev}, A.~Y., {Fabrika}, S., {et~al.} 2010, \mnras,
  405, 520

\bibitem[{{Habing}(1996)}]{habing96}
{Habing}, H.~J. 1996, \aapr, 7, 97

\bibitem[{{Habing} \& {Olofsson}(2003)}]{habing03}
{Habing}, H.~J. \& {Olofsson}, H., eds. 2003, {Asymptotic giant branch stars}

\bibitem[{{Habing} {et~al.}(2006){Habing}, {Sevenster}, {Messineo}, {van de
  Ven}, \& {Kuijken}}]{habing06}
{Habing}, H.~J., {Sevenster}, M.~N., {Messineo}, M., {van de Ven}, G., \&
  {Kuijken}, K. 2006, \aap, 458, 151

\bibitem[{{Hadfield} {et~al.}(2007){Hadfield}, {van Dyk}, {Morris}, {Smith},
  {Marston}, \& {Peterson}}]{hadfield07}
{Hadfield}, L.~J., {van Dyk}, S.~D., {Morris}, P.~W., {et~al.} 2007, \mnras,
  376, 248

\bibitem[{{Harwit} {et~al.}(2001){Harwit}, {Malfait}, {Decin}, {Waelkens},
  {Feuchtgruber}, \& {Melnick}}]{harwit01}
{Harwit}, M., {Malfait}, K., {Decin}, L., {et~al.} 2001, \apj, 557, 844

\bibitem[{{Homeier} {et~al.}(2003){Homeier}, {Blum}, {Pasquali}, {Conti}, \&
  {Damineli}}]{homeier03}
{Homeier}, N.~L., {Blum}, R.~D., {Pasquali}, A., {Conti}, P.~S., \& {Damineli},
  A. 2003, \aap, 408, 153

\bibitem[{{Imai} {et~al.}(2002){Imai}, {Deguchi}, {Fujii}, {Glass}, {Ita},
  {Izumiura}, {Kameya}, {Miyazaki}, {Nakada}, \& {Nakashima}}]{imai02}
{Imai}, H., {Deguchi}, S., {Fujii}, T., {et~al.} 2002, \pasj, 54, L19

\bibitem[{{Indebetouw} {et~al.}(2005){Indebetouw}, {Mathis}, {Babler}, {Meade},
  {Watson}, {Whitney}, {Wolff}, {Wolfire}, {Cohen}, {Bania}, {Benjamin},
  {Clemens}, {Dickey}, {Jackson}, {Kobulnicky}, {Marston}, {Mercer},
  {Stauffer}, {Stolovy}, \& {Churchwell}}]{indebetouw05}
{Indebetouw}, R., {Mathis}, J.~S., {Babler}, B.~L., {et~al.} 2005, \apj, 619,
  931

\bibitem[{{Josselin} {et~al.}(2000){Josselin}, {Blommaert}, {Groenewegen},
  {Omont}, \& {Li}}]{josselin00}
{Josselin}, E., {Blommaert}, J.~A.~D.~L., {Groenewegen}, M.~A.~T., {Omont}, A.,
  \& {Li}, F.~L. 2000, \aap, 357, 225

\bibitem[{{Kemper} {et~al.}(2001){Kemper}, {Waters}, {de Koter}, \&
  {Tielens}}]{kemper01}
{Kemper}, F., {Waters}, L.~B.~F.~M., {de Koter}, A., \& {Tielens}, A.~G.~G.~M.
  2001, \aap, 369, 132

\bibitem[{{Koornneef}(1983)}]{koornneef83}
{Koornneef}, J. 1983, \aap, 128, 84

\bibitem[{{Lamers} {et~al.}(2001){Lamers}, {Nota}, {Panagia}, {Smith}, \&
  {Langer}}]{lamers01}
{Lamers}, H.~J.~G.~L.~M., {Nota}, A., {Panagia}, N., {Smith}, L.~J., \&
  {Langer}, N. 2001, \apj, 551, 764

\bibitem[{{Landini} {et~al.}(1984){Landini}, {Natta}, {Salinari}, {Oliva}, \&
  {Moorwood}}]{landini84}
{Landini}, M., {Natta}, A., {Salinari}, P., {Oliva}, E., \& {Moorwood},
  A.~F.~M. 1984, \aap, 134, 284

\bibitem[{{Levesque} {et~al.}(2005){Levesque}, {Massey}, {Olsen}, {Plez},
  {Josselin}, {Maeder}, \& {Meynet}}]{levesque05}
{Levesque}, E.~M., {Massey}, P., {Olsen}, K.~A.~G., {et~al.} 2005, \apj, 628,
  973

\bibitem[{{Lucas} {et~al.}(2008){Lucas}, {Hoare}, {Longmore}, {Schr{\"o}der},
  {Davis}, {Adamson}, {Bandyopadhyay}, {de Grijs}, {Smith}, {Gosling},
  {Mitchison}, {G{\'a}sp{\'a}r}, {Coe}, {Tamura}, {Parker}, {Irwin}, {Hambly},
  {Bryant}, {Collins}, {Cross}, {Evans}, {Gonzalez-Solares}, {Hodgkin},
  {Lewis}, {Read}, {Riello}, {Sutorius}, {Lawrence}, {Drew}, {Dye}, \&
  {Thompson}}]{lucas08}
{Lucas}, P.~W., {Hoare}, M.~G., {Longmore}, A., {et~al.} 2008, \mnras, 391, 136

\bibitem[{{Martins} {et~al.}(2007){Martins}, {Genzel}, {Hillier}, {Eisenhauer},
  {Paumard}, {Gillessen}, {Ott}, \& {Trippe}}]{martins07}
{Martins}, F., {Genzel}, R., {Hillier}, D.~J., {et~al.} 2007, \aap, 468, 233

\bibitem[{{Matsuura} {et~al.}(2002){Matsuura}, {Yamamura}, {Cami}, {Onaka}, \&
  {Murakami}}]{matsuura02}
{Matsuura}, M., {Yamamura}, I., {Cami}, J., {Onaka}, T., \& {Murakami}, H.
  2002, \aap, 383, 972

\bibitem[{{Mauerhan} {et~al.}(2010){Mauerhan}, {Morris}, {Cotera}, {Dong},
  {Wang}, {Stolovy}, {Lang}, \& {Glass}}]{mauerhan10}
{Mauerhan}, J.~C., {Morris}, M.~R., {Cotera}, A., {et~al.} 2010, \apjl, 713,
  L33

\bibitem[{{Mauerhan} {et~al.}(2011){Mauerhan}, {Van Dyk}, \&
  {Morris}}]{mauerhan11}
{Mauerhan}, J.~C., {Van Dyk}, S.~D., \& {Morris}, P.~W. 2011, \aj, 142, 40

\bibitem[{{Mengel} \& {Tacconi-Garman}(2007)}]{mengel07}
{Mengel}, S. \& {Tacconi-Garman}, L.~E. 2007, \aap, 466, 151

\bibitem[{{Mermilliod} {et~al.}(2008){Mermilliod}, {Mayor}, \&
  {Udry}}]{mermilliod08}
{Mermilliod}, J.~C., {Mayor}, M., \& {Udry}, S. 2008, \aap, 485, 303

\bibitem[{{Messineo}(2004)}]{messineothesis}
{Messineo}, M. 2004, PhD thesis, Leiden Observatory, Leiden University,
  P.O.~Box 9513, 2300 RA Leiden, The Netherlands

\bibitem[{{Messineo} {et~al.}(2011){Messineo}, {Davies}, {Figer}, {Kudritzki},
  {Valenti}, {Trombley}, {Najarro}, \& {Rich}}]{messineo11}
{Messineo}, M., {Davies}, B., {Figer}, D.~F., {et~al.} 2011, \apj, 733, 41

\bibitem[{{Messineo} {et~al.}(2009){Messineo}, {Davies}, {Ivanov}, {Figer},
  {Schuller}, {Habing}, {Menten}, \& {Petr-Gotzens}}]{messineo09}
{Messineo}, M., {Davies}, B., {Ivanov}, V.~D., {et~al.} 2009, \apj, 697, 701

\bibitem[{{Messineo} {et~al.}(2010){Messineo}, {Figer}, {Davies}, {Kudritzki},
  {Rich}, {MacKenty}, \& {Trombley}}]{messineo10}
{Messineo}, M., {Figer}, D.~F., {Davies}, B., {et~al.} 2010, \apj, 708, 1241

\bibitem[{{Messineo} {et~al.}(2004){Messineo}, {Habing}, {Menten}, {Omont}, \&
  {Sjouwerman}}]{messineo04}
{Messineo}, M., {Habing}, H.~J., {Menten}, K.~M., {Omont}, A., \& {Sjouwerman},
  L.~O. 2004, \aap, 418, 103

\bibitem[{{Messineo} {et~al.}(2005){Messineo}, {Habing}, {Menten}, {Omont},
  {Sjouwerman}, \& {Bertoldi}}]{messineo05}
{Messineo}, M., {Habing}, H.~J., {Menten}, K.~M., {et~al.} 2005, \aap, 435, 575

\bibitem[{{Messineo} {et~al.}(2002){Messineo}, {Habing}, {Sjouwerman}, {Omont},
  \& {Menten}}]{messineo02}
{Messineo}, M., {Habing}, H.~J., {Sjouwerman}, L.~O., {Omont}, A., \& {Menten},
  K.~M. 2002, \aap, 393, 115

\bibitem[{{Minniti} {et~al.}(2010){Minniti}, {Lucas}, {Emerson}, {Saito},
  {Hempel}, {Pietrukowicz}, {Ahumada}, {Alonso}, {Alonso-Garcia}, {Arias},
  {Bandyopadhyay}, {Barb{\'a}}, {Barbuy}, {Bedin}, {Bica}, {Borissova},
  {Bronfman}, {Carraro}, {Catelan}, {Clari{\'a}}, {Cross}, {de Grijs},
  {D{\'e}k{\'a}ny}, {Drew}, {Fari{\~n}a}, {Feinstein}, {Fern{\'a}ndez
  Laj{\'u}s}, {Gamen}, {Geisler}, {Gieren}, {Goldman}, {Gonzalez}, {Gunthardt},
  {Gurovich}, {Hambly}, {Irwin}, {Ivanov}, {Jord{\'a}n}, {Kerins}, {Kinemuchi},
  {Kurtev}, {L{\'o}pez-Corredoira}, {Maccarone}, {Masetti}, {Merlo},
  {Messineo}, {Mirabel}, {Monaco}, {Morelli}, {Padilla}, {Palma}, {Parisi},
  {Pignata}, {Rejkuba}, {Roman-Lopes}, {Sale}, {Schreiber}, {Schr{\"o}der},
  {Smith}, {Sodr{\'e}}, {Soto}, {Tamura}, {Tappert}, {Thompson}, {Toledo},
  {Zoccali}, \& {Pietrzynski}}]{vista10}
{Minniti}, D., {Lucas}, P.~W., {Emerson}, J.~P., {et~al.} 2010, \na, 15, 433

\bibitem[{{Negueruela} {et~al.}(2011){Negueruela},
  {Gonz{\'a}lez-Fern{\'a}ndez}, {Marco}, \& {Clark}}]{negueruela11}
{Negueruela}, I., {Gonz{\'a}lez-Fern{\'a}ndez}, C., {Marco}, A., \& {Clark},
  J.~S. 2011, \aap, 528, A59

\bibitem[{{Negueruela} {et~al.}(2010){Negueruela},
  {Gonz{\'a}lez-Fern{\'a}ndez}, {Marco}, {Clark}, \&
  {Mart{\'{\i}}nez-N{\'u}{\~n}ez}}]{negueruela10}
{Negueruela}, I., {Gonz{\'a}lez-Fern{\'a}ndez}, C., {Marco}, A., {Clark},
  J.~S., \& {Mart{\'{\i}}nez-N{\'u}{\~n}ez}, S. 2010, \aap, 513, A74

\bibitem[{{Negueruela} \& {Schurch}(2007)}]{negueruela07}
{Negueruela}, I. \& {Schurch}, M.~P.~E. 2007, \aap, 461, 631

\bibitem[{{Nishiyama} {et~al.}(2006){Nishiyama}, {Nagata}, {Kusakabe},
  {Matsunaga}, {Naoi}, {Kato}, {Nagashima}, {Sugitani}, {Tamura}, {Tanab{\'e}},
  \& {Sato}}]{nishiyama06}
{Nishiyama}, S., {Nagata}, T., {Kusakabe}, N., {et~al.} 2006, \apj, 638, 839

\bibitem[{{Norci} {et~al.}(2002){Norci}, {Polcaro}, {Viotti}, \&
  {Rossi}}]{norci02}
{Norci}, L., {Polcaro}, V.~F., {Viotti}, R.~F., \& {Rossi}, C. 2002, \rmxaa,
  38, 83

\bibitem[{{Nota} {et~al.}(1995){Nota}, {Livio}, {Clampin}, \&
  {Schulte-Ladbeck}}]{nota95}
{Nota}, A., {Livio}, M., {Clampin}, M., \& {Schulte-Ladbeck}, R. 1995, \apj,
  448, 788

\bibitem[{{Olivier} {et~al.}(2001){Olivier}, {Whitelock}, \&
  {Marang}}]{olivier01}
{Olivier}, E.~A., {Whitelock}, P., \& {Marang}, F. 2001, \mnras, 326, 490

\bibitem[{{Ortiz} {et~al.}(2002){Ortiz}, {Blommaert}, {Copet}, {Ganesh},
  {Habing}, {Messineo}, {Omont}, {Schultheis}, \& {Schuller}}]{ortiz02}
{Ortiz}, R., {Blommaert}, J.~A.~D.~L., {Copet}, E., {et~al.} 2002, \aap, 388,
  279

\bibitem[{{Pasquali} {et~al.}(2002){Pasquali}, {Nota}, {Smith}, {Akiyama},
  {Messineo}, \& {Clampin}}]{pasquali02}
{Pasquali}, A., {Nota}, A., {Smith}, L.~J., {et~al.} 2002, \aj, 124, 1625

\bibitem[{{Peeters} {et~al.}(2002){Peeters}, {Hony}, {Van Kerckhoven},
  {Tielens}, {Allamandola}, {Hudgins}, \& {Bauschlicher}}]{peeters02}
{Peeters}, E., {Hony}, S., {Van Kerckhoven}, C., {et~al.} 2002, \aap, 390, 1089

\bibitem[{{Pierce} {et~al.}(2000){Pierce}, {Jurcevic}, \&
  {Crabtree}}]{pierce00}
{Pierce}, M.~J., {Jurcevic}, J.~S., \& {Crabtree}, D. 2000, \mnras, 313, 271

\bibitem[{{Price} {et~al.}(2001){Price}, {Egan}, {Carey}, {Mizuno}, \&
  {Kuchar}}]{price01}
{Price}, S.~D., {Egan}, M.~P., {Carey}, S.~J., {Mizuno}, D.~R., \& {Kuchar},
  T.~A. 2001, \aj, 121, 2819

\bibitem[{{Rieke} {et~al.}(1985){Rieke}, {Lebofsky}, \& {Low}}]{rieke85}
{Rieke}, G.~H., {Lebofsky}, M.~J., \& {Low}, F.~J. 1985, \aj, 90, 900

\bibitem[{{Robitaille} {et~al.}(2007){Robitaille}, {Cohen}, {Whitney}, {Meade},
  {Babler}, {Indebetouw}, \& {Churchwell}}]{robitaille07}
{Robitaille}, T.~P., {Cohen}, M., {Whitney}, B.~A., {et~al.} 2007, \aj, 134,
  2099

\bibitem[{{Schuller} {et~al.}(2003){Schuller}, {Ganesh}, {Messineo}, {Moneti},
  {Blommaert}, {Alard}, {Aracil}, {Miville-Desch{\^e}nes}, {Omont},
  {Schultheis}, {Simon}, {Soive}, \& {Testi}}]{schuller03}
{Schuller}, F., {Ganesh}, S., {Messineo}, M., {et~al.} 2003, \aap, 403, 955

\bibitem[{{Schultheis} \& {Glass}(2001)}]{schultheis01}
{Schultheis}, M. \& {Glass}, I.~S. 2001, \mnras, 327, 1193

\bibitem[{{Sevenster}(1999)}]{sevenster99}
{Sevenster}, M.~N. 1999, \mnras, 310, 629

\bibitem[{{Sevenster}(2002)}]{sevenster02}
{Sevenster}, M.~N. 2002, \aj, 123, 2772

\bibitem[{{Shara} {et~al.}(2011){Shara}, {Faherty}, {Zurek}, {Moffat}, {Gerke},
  {Doyon}, {Artigau}, \& {Drissen}}]{shara11}
{Shara}, M.~M., {Faherty}, J.~K., {Zurek}, D., {et~al.} 2011, ArXiv e-prints

\bibitem[{{Sjouwerman} {et~al.}(2009){Sjouwerman}, {Capen}, \&
  {Claussen}}]{sjouwerman09}
{Sjouwerman}, L.~O., {Capen}, S.~M., \& {Claussen}, M.~J. 2009, \apj, 705, 1554

\bibitem[{{Sjouwerman} {et~al.}(1998){Sjouwerman}, {van Langevelde},
  {Winnberg}, \& {Habing}}]{sjouwerman98}
{Sjouwerman}, L.~O., {van Langevelde}, H.~J., {Winnberg}, A., \& {Habing},
  H.~J. 1998, \aaps, 128, 35

\bibitem[{{Skiff}(2010)}]{skiff10}
{Skiff}, B.~A. 2010, VizieR Online Data Catalog, 1, 2023

\bibitem[{{Sloan} {et~al.}(2003){Sloan}, {Kraemer}, {Price}, \&
  {Shipman}}]{sloan03}
{Sloan}, G.~C., {Kraemer}, K.~E., {Price}, S.~D., \& {Shipman}, R.~F. 2003,
  \apjs, 147, 379

\bibitem[{{Smith} {et~al.}(2004){Smith}, {Vink}, \& {de Koter}}]{smith04}
{Smith}, N., {Vink}, J.~S., \& {de Koter}, A. 2004, \apj, 615, 475

\bibitem[{{Speck} {et~al.}(2000){Speck}, {Barlow}, {Sylvester}, \&
  {Hofmeister}}]{speck00}
{Speck}, A.~K., {Barlow}, M.~J., {Sylvester}, R.~J., \& {Hofmeister}, A.~M.
  2000, \aaps, 146, 437

\bibitem[{{Spitzer Science}(2009)}]{spitzer09}
{Spitzer Science}, C. 2009, VizieR Online Data Catalog, 2293, 0

\bibitem[{{Stead} \& {Hoare}(2009)}]{stead09}
{Stead}, J.~J. \& {Hoare}, M.~G. 2009, \mnras, 400, 731

\bibitem[{{Sylvester} {et~al.}(1999){Sylvester}, {Kemper}, {Barlow}, {de Jong},
  {Waters}, {Tielens}, \& {Omont}}]{sylvester99}
{Sylvester}, R.~J., {Kemper}, F., {Barlow}, M.~J., {et~al.} 1999, \aap, 352,
  587

\bibitem[{{The}(2005)}]{denis}
{The}, C.~D. 2005, VizieR Online Data Catalog, 2263, 0

\bibitem[{{Tsuji} {et~al.}(1997){Tsuji}, {Ohnaka}, {Aoki}, \&
  {Yamamura}}]{tsuji97}
{Tsuji}, T., {Ohnaka}, K., {Aoki}, W., \& {Yamamura}, I. 1997, \aap, 320, L1

\bibitem[{{Umana} {et~al.}(2010){Umana}, {Buemi}, {Trigilio}, {Leto}, \&
  {Hora}}]{umana10}
{Umana}, G., {Buemi}, C.~S., {Trigilio}, C., {Leto}, P., \& {Hora}, J.~L. 2010,
  \apj, 718, 1036

\bibitem[{{van der Hucht}(2001)}]{vanderhucht01}
{van der Hucht}, K.~A. 2001, VizieR Online Data Catalog, 3215, 0

\bibitem[{{van der Veen} \& {Habing}(1988)}]{vanderveen88}
{van der Veen}, W.~E.~C.~J. \& {Habing}, H.~J. 1988, \aap, 194, 125

\bibitem[{{van Loon} {et~al.}(2003){van Loon}, {Gilmore}, {Omont}, {Blommaert},
  {Glass}, {Messineo}, {Schuller}, {Schultheis}, {Yamamura}, \&
  {Zhao}}]{vanloon03}
{van Loon}, J.~T., {Gilmore}, G.~F., {Omont}, A., {et~al.} 2003, \mnras, 338,
  857

\bibitem[{{Vauterin} \& {Dejonghe}(1998)}]{vauterin98}
{Vauterin}, P. \& {Dejonghe}, H. 1998, \apj, 500, 233

\bibitem[{{Verheyen} {et~al.}(2011){Verheyen}, {Menten}, \& {
  Messineo}}]{verheyen11}
{Verheyen}, L., {Menten}, K.~M., \& { Messineo}, M. 2011, \aap\ to be submitted

\bibitem[{{Verhoelst} {et~al.}(2009){Verhoelst}, {van der Zypen}, {Hony},
  {Decin}, {Cami}, \& {Eriksson}}]{verhoelst09}
{Verhoelst}, T., {van der Zypen}, N., {Hony}, S., {et~al.} 2009, \aap, 498, 127

\bibitem[{{Waters}(2010)}]{waters10}
{Waters}, L.~B.~F.~M. 2010, in Astronomical Society of the Pacific Conference
  Series, Vol. 425, Hot and Cool: Bridging Gaps in Massive Star Evolution, ed.
  {C.~Leitherer, P.~Bennett, P.~Morris, \& J.~van Loon}, 267

\bibitem[{{Whitelock} \& {Feast}(1994)}]{whitelock94}
{Whitelock}, P. \& {Feast}, M. 1994, \apss, 217, 153

\bibitem[{{Whitelock} {et~al.}(2000){Whitelock}, {Marang}, \&
  {Feast}}]{whitelock00}
{Whitelock}, P., {Marang}, F., \& {Feast}, M. 2000, \mnras, 319, 728

\bibitem[{{Whitelock} {et~al.}(2003){Whitelock}, {Feast}, {van Loon}, \&
  {Zijlstra}}]{whitelock03}
{Whitelock}, P.~A., {Feast}, M.~W., {van Loon}, J.~T., \& {Zijlstra}, A.~A.
  2003, \mnras, 342, 86

\bibitem[{{Williams} {et~al.}(1994){Williams}, {van der Hucht}, {Kidger},
  {Geballe}, \& {Bouchet}}]{williams94}
{Williams}, P.~M., {van der Hucht}, K.~A., {Kidger}, M.~R., {Geballe}, T.~R.,
  \& {Bouchet}, P. 1994, \mnras, 266, 247

\bibitem[{{Wright} \& {Barlow}(1975)}]{wright75}
{Wright}, A.~E. \& {Barlow}, M.~J. 1975, \mnras, 170, 41

\bibitem[{{Yang} \& {Jiang}(2011)}]{yang11}
{Yang}, M. \& {Jiang}, B.~W. 2011, \apj, 727, 53

\bibitem[{{Zacharias} {et~al.}(2004){Zacharias}, {Monet}, {Levine}, {Urban},
  {Gaume}, \& {Wycoff}}]{zacharias04}
{Zacharias}, N., {Monet}, D.~G., {Levine}, S.~E., {et~al.} 2004, in Bulletin of
  the American Astronomical Society, Vol.~36, American Astronomical Society
  Meeting Abstracts, 1418

\end{thebibliography}

\end{document}